\documentclass[aps,pra,amsmath,notitlepage,twocolumn,superscriptaddress,nofootinbib,eprints]{revtex4-2} 
\pdfoutput=1
\usepackage[utf8]{inputenc}
\usepackage[english]{babel}
\usepackage[T1]{fontenc}
\usepackage{microtype}

\usepackage{amsmath,amsfonts,amssymb,amsthm,bm,bbm,bbold}
\usepackage{braket}
\usepackage[dvipsnames]{xcolor}
\usepackage{mathpazo}
\usepackage[normalem]{ulem}

\usepackage{pifont}

\usepackage[breaklinks=true,colorlinks=true,linkcolor=teal,urlcolor=teal,citecolor=teal]{hyperref}

\newcommand{\be}{\begin{equation}}
\newcommand{\ee}{\end{equation}}
\newcommand{\bae}{\begin{eqnarray}} \newcommand{\eae}{\end{eqnarray}}

\def\EE{\mathbbm{E}}
\def\id{\mathbbm{1}}
\def\Tr{\hbox{Tr}}
\def\fb{\text{fb}}
\def\sigmaCM{\boldsymbol{\sigma}}

\begin{document}

\title{A pedagogical introduction to continuously monitored quantum systems and measurement-based feedback}
\author{Francesco Albarelli}
\email{francesco.albarelli@gmail.com}
\affiliation{Scuola Normale Superiore, I-56126 Pisa, Italy}

\author{Marco G. Genoni}
\email{marco.genoni@fisica.unimi.it}
\affiliation{Dipartimento di Fisica ``Aldo Pontremoli'', Università degli Studi di Milano, I-20133 Milan, Italy}

\begin{abstract}
In this manuscript we present a pedagogical introduction to continuously monitored quantum systems.
We start by giving a simplified derivation of the Markovian master equation in Lindblad form, in the spirit of collision models and input-output theory, which describes the unconditional dynamics of a continuously monitored system.
The same formalism is then exploited to derive stochastic master equations that describe the conditional dynamics.
We focus on the two most paradigmatic examples of continuous monitoring: continuous photodetection, leading to a discontinuous dynamics with ``quantum jumps'', and continuous homodyne measurements, leading to a diffusive dynamics.
We then present a derivation of feedback master equations that describe the dynamics (either conditional or unconditional) when the continuous measurement photocurrents are fed back to the system as a linear driving Hamiltonian, a paradigm known as linear Markovian feedback.
In the second part of the manuscript we focus on continuous-variable Gaussian systems: we first present the equations for first and second moments describing the dynamics under continuous general-dyne measurements, and we then discuss in more detail the conditional and unconditional dynamics under Markovian and state-based feedback.
\end{abstract}

\maketitle

\section{Introduction}

The act of measuring a quantum system, and thus extracting classical information from it, has a special place in the theory of quantum mechanics.
While in standard textbook treatments measurement is often assumed to be an instantaneous and maximally invasive process, more general situations are possible.
In particular, by indirectly measuring the main system through an auxiliary system in interaction with it, it is possible to implement so-called weak measurements.
These are not fully disruptive, but at the same time they extract little information on the system's state.
This paper deals with situations in which it is possible to \emph{weakly} measure the system \emph{continuously} in time, i.e. to \emph{continuously monitor} a quantum system, leading to stochastic conditional dynamics for the quantum state, typically referred to as \emph{quantum trajectories}.
Single quantum trajectories of continuously monitored quantum systems has been now experimentally shown in different platforms, such as superconducting circuits~\cite{Murch2013,Campagne-Ibarcq2016,Ficheux2017,Minev2019}, optomechanical~\cite{Wieczorek2015,Rossi2018,Mason2018} and hybrid~\cite{Thomas2020} quantum systems.
Recently, continuous feedback protocols able to cool mechanical oscillators towards their quantum ground state have also been demonstrated~\cite{Rossi2018a,Magrini2021,Tebbenjohanns2021}.

There is no shortage of introductory material and textbooks on the topic of continuously monitored quantum systems.
Indeed, most of the topics covered in this manuscript are can also be found (possibly presented with different approaches) in Refs.~\cite{Wiseman1994,Jacobs2006,wiseman2010quantum,Jacobs2014a,Serafini2023,Rouchon2022}.
Nonetheless, we believe that our presentation and choice of subjects has no exact equivalent in a single introductory paper, and may be useful for students and newcomers to this field. 

Our approach in the derivations is to favour intuition and physical insight, without pursuing formal and mathematical rigour.
Moreover, the approach we follow hinges conceptually on a (more or less formal) discretization of the continuous-time evolution.
This connection can be made more precise in the context of the so-called collision models~\cite{Ciccarello2017,Ciccarello2022}.
As a matter of fact, the equations we derive in this manuscript can alternatively be obtained simply in term of qubit ancillas interacting with the sytem in lieu of the field~\cite{Brun2002,Gross2017}.
In particular, Ref.~\cite{Gross2017} and~\cite{Rouchon2022} are two references that we think complement well our treatment here.
Other useful and more comprehensive resources are the books~\cite{Carmichael1993,wiseman2010quantum,Jacobs2014a}.

Originally, the topic of continuous measurements of quantum systems and its connection to the theory of open quantum systems was studied from different perspectives, through a more mathematically oriented approach, see e.g. Refs.~\cite{Barchielli2006,Barchielli2013}.
Since continuous measurements can be used as a method to continuously track the system's state, there is a strong connection with the classical problem of filtering, i.e. determining the state of a dynamical system subject to stochastic noise from imperfect observations of the state.
Indeed, the equations that we call here stochastic master equations are also known as quantum filtering equations~\cite{Bouten2006,boutenDiscreteApproximationQuantum2008,Gough2018} dating back to the seminal work of Belavkin~\cite{Belavkin1995,belavkinMeasurementFilteringControl1999}.
We also remark how quantum trajectories have been introduced as a tool for more efficient numerical simulations of open quantum systems, leading to a method known as \emph{Monte Carlo wave function}~\cite{Dalibard1992,Dum1992}.
We finally mention that quantum trajectories have also appeared in the context of collapse models~\cite{Bassi2013a,Tilloy2020}, i.e. generalizations of quantum mechanics that aim to account for the absence of macroscopic superpositions.

As stated in the title, this manuscript is not intended as a review, but merely as a relatively simple pedagogical introduction to the fundamental concepts and equations of this field.
This is reflected also from the point of view of the bibliography: it is not meant to capture the myriad interesting results in this vast field.
When possible we cite other introductory material, such as tutorials and textbooks, having in mind accessibility for students and newcomers.

The manuscript is structured as follows.
In Section~\ref{sec:ME} we present a derivation of the Markovian master equation in the spirit of collision models and input-output theory.
In Section~\ref{sec:SME} we present a derivation the stochastic differential equation that dictates the conditional dynamics of the quantum system, when it is continuously monitored by performing measurements on the output field, either homodyne measurements or photodetection.
In Section~\ref{sec:feedback} we present a derivation of the equations that govern the dynamics of the system  (both conditional and unconditional) when the continuous measurement results are fed back to drive the system with a linear Markovian feedback.
In Section~\ref{sec:Gaussian} we present the dynamical equations for Gaussian systems in presence of Markovian and state-based feedback, this is intended as a complement to Ref.~\cite{Genoni2016} which contains all the prerequisite material.
Section~\ref{sec:outlooks} concludes the paper with some perspectives of current research topics within the field.

\section{Derivation of Markovian master equations}\label{sec:ME}
In this section we present a simple derivation of the celebrated Gorini-Kossakowski-Sudarshan-Lindblad master equation~\cite{breuer2002theory,Chruscinski2017a,Campaioli2023} (which we mostly call only Lindblad equation, for semantic convenience).
We do not give rigorous proofs, but we try to sketch the main ideas behind the derivation of the equations.
Note that this derivation is quite different from microscopic derivations in the weak-coupling approximation, e.g. Refs.~\cite[Sec.~3.3]{breuer2002theory} and \cite[Sec.~6.1]{Walls2011}, and closely related to derivations based on collision models~\cite{Ciccarello2017,Ciccarello2022}.

\subsection{Hamiltonian dynamics}
Let us briefly recall how unitary dynamics is physically generated by the Hamiltonian of the system, which is a self-adjoint operator with a spectrum bounded from below\footnote{We remark that in some cases we will work with \emph{effective} Hamiltonians that are not bounded from below.}.
The differential equation which dictates the dynamic of a pure state is the Schrödinger equation
\begin{equation}
\label{eq:schro}
\frac{ \mathrm{d} \ket{\psi(t)} } {\mathrm{d} t } =  - \frac{i}{\hbar} \hat{H}(t) \ket{\psi(t)} \,;
\end{equation}
this is equivalent to an equation for an observable $\hat{O}$ in the Heisenberg picture:
\begin{equation}
\label{eq:Heis}
\frac{ \mathrm{d} \hat O (t)}{\mathrm{d} t } =  \frac{i}{\hbar}  \left[  \hat H(t) , \hat O (t) \right] \,.
\end{equation}
Eq.~\eqref{eq:schro} in the Schrödinger picture can equivalently be rewritten as a Liouville-Von Neumann equation for the density operator
\begin{equation}
\label{eq:Liou-VN}
\frac{ \mathrm{d} \varrho(t) } {\mathrm{d} t } =  -\frac{i}{\hbar} \left[  \hat{H}(t) , \varrho (t) \right] \,,
\end{equation}
this equation has the same form as~\eqref{eq:Heis}, but with the opposite sign.
The unitary operator expressing the solution to equations~\eqref{eq:Liou-VN} and~\eqref{eq:schro} for any initial state can formally be written as
\begin{equation}
\hat{U}(t,t_0) = \mathcal{T} \exp\left[ \frac{-i}{\hbar} \int_{t_0}^{t} d t' \hat{H}(t') \right] \,,\label{eq:TdepU}
\end{equation}
the time-ordering operator $\mathcal{T}$ has the effect of putting non-commuting operators in the correct chronological order.
Formally we have
\begin{equation}
\mathcal{T} \left( \hat{H}(t_1) \hat{H}(t_2) \right) = \begin{cases}
\hat{H}(t_1) \hat{H}(t_2) \quad \text{if } \quad t_1 < t_2 \\
\hat{H}(t_2) \hat{H}(t_1) \quad \text{if } \quad t_1 > t_2 
\end{cases} \,
\end{equation}
and analogously for more than two operators; this definition is applied to the exponential~\eqref{eq:TdepU} by Taylor expansion. 
When the Hamiltonian is time independent, we do not need time ordering and the evolution is homogeneous in time (it is only described by the elapsed time $t$):
\begin{equation}
\hat{U}(t) = \exp\left[-\frac{i}{\hbar}  \hat{H} t \right] \,;
\end{equation}
note that from now on we always rescale units appropriately such that $\hbar = 1$.

Unitary evolutions are due only to the Hamiltonian of the system, therefore they describe \emph{closed} quantum systems, more general evolutions arise when the system interacts with an external quantum system.
Usually, the external system is conceived as an environment, i.e. a ``large'' quantum system with a Hilbert space of much higher dimensionality than the main system, that is not under experimental control.
In this situation we usually talk of \emph{open} quantum systems~\cite{breuer2002theory}.
The external system can also be conceived as auxiliary degrees of freedom that can be measured and exploited to extract information on the main system.
We will see that these two situations are intimately connected, and the distinction may be blurred at times.

The topic of this section is precisely the dynamics of open quantum systems, when we do not have any access to the degrees of freedom of the environment (unconditional evolution).
In the next section we will see how things change when we can actually access and measure them (conditional evolution).

\subsubsection{Lindblad master equation}
\label{subsec:Lindblad}
This derivation is obtained using input-output theory, in a way that is going to make the transition to stochastic master equations transparent.
In order to derive the input-output formalism we consider the following physical description (see for example Ref.~\cite{Lammers}): a generic quantum system\footnote{We are not making any assumption on the dimension of the Hilbert space associated to the quantum system $\mathcal{S}$; this can correspond, e.g., to a qubit or to a quantum harmonic oscillator.} $\mathcal{S}$ interacts with an environment (bath) $\mathcal{B}$ corresponding to a continuum of bosonic modes, such that the total Hamiltonian governing the evolution of system and bath reads
\begin{align}
\hat{H} = \hat{H}_\mathcal{S} + \hat{H}_\mathcal{B} + \hat{H}_{\sf int} \,,
\end{align}
where $\hat{H}_\mathcal{S}$ is, for the moment, left generic, while
\begin{align}
\hat{H}_\mathcal{B} = \int_0^\infty d\omega \, \omega \, \hat{b}^\dag_\omega \hat{b}_\omega \,,
\end{align}
with the operators $\hat{b}_\omega$ satisfying the commutation relation $[ \hat{b}_\omega, \hat{b}^\dag_{\omega^\prime}]=\delta(\omega - \omega^\prime)$.
We then assume that the interaction Hamiltonian is linear in the operators $\hat{b}_\omega$, 
\begin{align}
\hat{H}_{\sf int} = i \int_0^\infty d\omega \sqrt{\frac{\kappa(\omega)}{2\pi}} \left( \hat{c} \hat{b}^\dag_\omega - \hat{c}^\dag \hat{b}_\omega\right) \,,
\label{eq:Hint}
\end{align}
where $\hat{c}$ is a given system operator, and $\kappa(\omega)$ quantifies the strength of the interaction between the system and the bath mode at frequency $\omega$.

We now move to the interaction picture with respect to $\hat{H}_0 = \hat{H}_\mathcal{S} + \hat{H}_\mathcal{B}$ and make the following further assumptions.
\begin{itemize}
\item {\em First Markov approximation}: by considering an interaction strongly localized in space, we can assume that interaction strength is constant in a certain (large) frequency bandwidth, i.e. $\kappa(\omega) = \kappa$ for $\omega \in \mathcal{W} = [\omega_0 - \theta, \omega_0 + \theta]$ (where $\omega_0$ is the {\em characteristic frequency} of the system), and negligible outside this interval.
\item We assume that the interaction-picture system operator acquires a time-dependent phase: $\hat{c}(t) = \hat{c} e^{-i \omega_0(t-t_0)}$, where $t_0$ denotes the initial time, where Schrödinger and interaction picture coincide.
We also assume that the timescale set by the interaction is such that $\tau=1/\kappa \gg 1/\omega_0$, and that all other time-dependencies are assumed to be negligible on this timescale.

For example in atomic systems with free Hamiltonian $\omega_0 \hat{J}_z$, the ladder operator $\hat{c}=\hat{J}_{-} = \hat{J}_x - i \hat{J}_y$ satisfies this property, and the interaction Hamiltonian~\eqref{eq:Hint} is obtained after making the rotating wave approximation. Similarly one can consider a cavity mode with free Hamiltonian $\omega_0 \hat{a}^\dag \hat{a}$, and with $\hat{c}=\hat{a}$.
One can also end up with operators $\hat{c}$ that do not satisfy this property, e.g. $\hat{c}=\hat{x}$ the Hermitian position operator of a oscillator, but this is usually an effective coupling obtained, e.g., after an adiabatic elimination of the optical cavity~\cite{Doherty1999}.
\end{itemize}

By defining the so-called {\em input modes} $b_t$ as
\begin{align} 
\hat{b}_t := \frac{1}{\sqrt{2\pi}} \int_\mathcal{W} d\omega \, \hat{b}_\omega e^{-i (\omega - \omega_0)(t-t_0)} 
\label{eq:bt}
\end{align}
the time-dependent interaction Hamiltonian (in interaction picture) can be written as
\begin{align}
\hat{H}_{\sf int}(t) = i \sqrt{\kappa}\left( \hat{c} \hat{b}^\dag_t - \hat{c}^\dag \hat{b}_t \right) \,. \label{eq:input_int_H}
\end{align}
We now want to evaluate the commutation relation for the input modes $\hat{b}_t$, finding
\begin{align}
& \left[ \hat{b}_t , \hat{b}_{t'}^\dag \right] = \\
& = \frac{1}{2\pi} \iint_{\omega_0 -\theta}^{\omega_0+\theta} d\omega
 d\omega^\prime e^{-i (\omega - \omega_0)(t-t_0)} e^{i (\omega^\prime - \omega_0)(t'-t_0)} [\hat{b}_\omega, \hat{b}_{\omega'}] \nonumber \\
&= \frac{1}{2\pi} \int_{\omega_0 -\theta}^{\omega_0+\theta} d\omega \, e^{-i (\omega - \omega_0)(t-t')} = \frac{1}{2\pi} \int_{-\theta}^{\theta} d\bar\omega \, e^{-i \bar\omega(t-t')} \nonumber \,
\end{align}
where we have exploited the commutation relation for the operators $\hat{b}_\omega$, and where we have changed integration variable to $\bar{\omega} = \omega - \omega_0$. If we now assume that the frequency bandwidth is much larger than all the other frequencies entering into the dynamics (i.e. $\theta \gg \kappa$), the integral can be extended to the whole real axis $\mathbbm{R}$\footnote{One should notice that, in order to properly define the frequency interval $\mathcal{W}$, one always needs that $\omega_0 - \theta >0$. As a consequence, the condition $\theta \gg \kappa$ also implies $\omega_0 \gg \kappa$, that is that the characteristic frequency of the system is much larger than all the other frequencies in the dynamics.
We also remark that these conditions will have to hold with respect to all other frequencies that may appear in the system Hamiltonian $\hat{H}_S$ and that we are neglecting in this simplified derivation of the master equation.}. 
As a consequence one obtains that the input modes satisfy the commutation relation
\begin{equation}
\label{eq:input_comm_delta}
\left[ \hat{b}_t , \hat{b}_{t'}^\dag \right] = \delta(t-t') \,.
\end{equation}
These are also called ``white noise'' operators, as their commutation relations implies that they are delta-correlated in time, exactly like classical white noise.

The intuition behind this Markovian input-output theory is that the environment can be thought as an infinite collection of uncorrelated quantum systems, all in the same state and that the main system interacts at each time with a different system from this collection.
This is the continuum limit of so-called collision models~\cite{Ciccarello2017,Ciccarello2022}.
In this sense, we made the typographic choice of denoting the input modes as $\hat{b}_t$ and not as $\hat{b}(t)$ to highlight the fact that $t$ is a label identifying a particular input operator in the collection $\{\hat{b}_t\}_{t \in \mathbbm{R}}$, and not a variable.
In a quantum optical setting the input operators $\hat{b}_t$ represent travelling light fields, while the system (usually an optical cavity) is fixed and localized and interacts with the travelling light impinging on it.

We can see that the commutator~\eqref{eq:input_comm_delta} at equal times is not a well defined quantity, so we heuristically go around this problem by integrating the input operators as
\begin{equation}
\label{eq:int_bin}
\delta \hat{b}_t = \int_{t}^{t+\delta t} \hat{b}_{t'} \, dt' \, ,
\end{equation}
which implies
\begin{equation}
\left[ \delta \hat{b}_{t}, \delta \hat{b}_{t}^\dag\right] = \id \delta t \,.
\end{equation}
For an infinitesimal $\delta t \to dt$ we have that $\delta \hat{b}_t \to d\hat{b}_t = \hat{b}_t  d t $; this leads to the following identity
\begin{equation}
\left[d\hat{b}_t , d\hat{b}_t^\dag \right] = \left[ \hat{b}_{t}, \hat{b}_{t}^\dag\right] dt^2 = \id dt \,.
\end{equation}
By following the approach pursued in Ref.~\cite{Wiseman1994}, we now define a new set of bosonic operators $\hat{B}_{t}$ satisfying the relation $\hat{b}_t dt = \hat{B}_{t} \sqrt{dt}$
\footnote{In more mathematical literature the operator $d \hat{b}_\text{in} \equiv \hat{B}_{t} \sqrt{dt}$ is called a quantum Wiener increment, and these concepts belong to the field known as quantum stochastic calculus~\cite{Parthasarathy1992}.}, these operators are proper creation and annihilation operator at each time, i.e.
\begin{equation}
\left[ \hat{B}_{t}, \hat{B}_{t}^\dag \right] = \id \,. \label{eq:Bcommutation}
\end{equation}
As $\hat{B}_t$ corresponds to annihilation operators of {\em proper} quantum harmonic oscillators, we can now formally specify that the global Hilbert space of corresponds to 
\begin{align}
\mathbbm{H} = \mathbbm{H}_{\mathcal{S}} \bigotimes_{t \in \mathbbm{R}} \mathbbm{H}_{\mathcal{B}_t}
\end{align}
that is, to the system Hilbert space $\mathbbm{H}_{\mathcal{S}}$, tensor product with an (uncountable) infinity of infinite-dimensional Hilbert spaces $\mathbbm{H}_{\mathcal{B}_t}$, each one corresponding to a quantum harmonic oscillator, labelled by the time variable $t$.

In order to study the evolution of the system quantum state $\varrho(t)$ we now follow these hypothesis:
\begin{itemize}
\item 
{\em Born-Markov approximation}: at each time $t$, the global state of system and environment $R(t)$ is a factorized state with the same reduced state for the environment, i.e.
\begin{align}
R(t) = \varrho(t) \otimes \mu_t
\end{align} 
In the following we will always assume that the state of each mode is the vacuum,  $\mu_t = |0\rangle_t\langle 0|$, satisfying $\hat B_t \ket{0}_t = 0 $ and $\hat B_t^\dag \ket{0}_t = \ket{1}_t$ (a single photon state) at each $t$ (we should remark that we can properly define vacuum and Fock states as $\hat{B}_t$ is a proper annihilation operator satisfying (\ref{eq:Bcommutation})).
This choice, in terms of the original operators $\{\hat{b}_t \}$, implies that the expectation value of the anti-commutator is again a Dirac delta
\begin{equation}
\left\langle \left\{ \hat{b}_t , \hat{b}_{t'}^\dag \right\} \right\rangle = \delta(t-t') \,. \label{eq:zerotemperature}
\end{equation}
Physically, this corresponds to assuming that the characteristic frequency of the system $\omega_0$ is large enough so that the environment can be assumed to be at zero-temperature.
More precisely, it is a condition on the average number of thermal photons: $N( \omega_0 ) = 1 / ( e^{\hbar \omega_0 / ( k_B T )} -1) \ll 1$, valid, e.g., for optical frequencies at room temperature.
In Appendix~\ref{s:genericbath} we relax this assumption and describe the evolution corresponding to thermal and squeezed environments, as well as in the case of a coherent driving.

The conditions in Eqs.~\eqref{eq:zerotemperature} and~\eqref{eq:input_comm_delta} represent our Markovianity assumptions.
The physical meaning is that the environment subsystems interacting at different times are completely uncorrelated and the interaction itself is instantaneous. This is clearly an idealisation which relies on a separation of timescales between the dynamics of the system and the environment.
Roughly speaking, we are assuming that the correlation time of the environment is much shorter than the timescales governing the evolution of the system\footnote{
Alternatively, we could say that in this approximation $dt$ is mathematically an infinitesimal quantity, but physically corresponds to the smallest timescale of the system and anything happening on shorter timescales can be disregarded.}.
\item 
The global state evolves in a infinitesimal time $dt$ as
\begin{align}
R(t+dt) = \hat{U}(t,t+dt) R(t) \hat{U}(t,t+dt)^\dag \,
\end{align}
where $\hat{U}(t,t+ dt) = e^{-i \hat{H}_\text{int} (t) \, dt}$.
\item
The state of the system at time $t+dt$ is obtained by tracing out the degrees of freedom of the environment, i.e.
\begin{align}
\varrho(t+dt) = \Tr_\mathcal{B} [ R(t+dt) ] \,,
\end{align}
where here $\Tr_\mathcal{B}[\bullet]$ denotes the partial trace on the Hilbert space corresponding to the oscillator mode $\hat{B}_t$.
\end{itemize}

We can now write the unitary evolution operator $\hat{U}(t,t+dt)$ as a function of the operators $\hat{B}_{t}$ as follows
\begin{equation}
\begin{split}
\hat{U}(t,t+ dt) &= e^{-i \hat{H}_\text{int} (t) \, dt}\\
&= \exp\left[ \sqrt{\kappa} dt \left( \hat{c} \otimes \hat{b}_{t}^\dag- \hat{c}^\dag \otimes \hat{b}_t  \right) \right] \\
&= \exp\left[ \sqrt{\kappa \, dt} \left( \hat{c} \otimes \hat{B}_{t}^\dag  - \hat{c}^\dag \otimes \hat{B}_{t}\right)\right] \label{eq:input_U_sqrtdt};
\end{split}
\end{equation}
The key feature of the form~\eqref{eq:input_U_sqrtdt} is that it is an exponential of the dimensionless variable $\sqrt{\kappa dt}$.
Thus, to obtain terms up to first order in $dt$, we need to expand the exponential up to second order as follows
\begin{equation}
\label{eq:input_U_series}
\begin{split}
&\hat U(t,t+ dt) \approx  \id \otimes \id + \left( \hat c \otimes \hat{B}_{t}^\dag - \hat c^\dag \otimes \hat{B}_{t} \right) \sqrt{\kappa dt} + \\
 & + \frac{1}{2}\left( \hat c^{\dag 2} \otimes \hat{B}_t^2  + \hat c^2 \otimes \hat{B}_t^{\dag 2} - \hat c \hat c^\dag \otimes \hat{B}_t^{\dag} \hat{B}_t - \hat c^\dag \hat c \otimes \hat{B}_t \hat{B}_t^{\dag}   \right) \kappa dt ;
\end{split}
\end{equation}
we note here that sometimes we are going to include the parameter $\kappa$ in the definition of the operator $\hat c$ by rescaling it as $\hat c \to \sqrt{\kappa} \hat c$.

Now we bring into the calculation the assumptions above, and we compute the evolution of the density matrix of the system as
\begin{equation}
\varrho(t+dt) = \Tr_\mathcal{B} \left[ \hat U(t,t+dt) \left( \varrho(t) \otimes \ket{0}_t \! \bra{0} \right) \hat U(t,t+dt)^\dag\ \right].
\end{equation}
By inserting the expansion~\eqref{eq:input_U_series} we get
\begin{align}
R(t+dt) &= \varrho(t) \otimes |0\rangle\langle 0| +  \label{eq:globalev} \\
& + \sqrt{\kappa} \left( \hat c \varrho(t) \otimes |1\rangle \langle 0| + \varrho(t)  \hat c^\dag \otimes |0\rangle \langle 1| \right) \, \sqrt{dt}  \nonumber \\
& + \kappa \left( \hat c \varrho(t) \hat c^\dag \otimes |1\rangle \langle 1| \right) \, dt + \nonumber \\
& + \frac{\kappa}{2} ( \sqrt{2} \hat c^2 \varrho(t) \otimes |2\rangle\langle 0| - \hat{c}^\dag \hat c \varrho(t) \otimes |0\rangle\langle 0 | + \mathrm{h.c.}) \, dt \,. \nonumber
\end{align}
By explicitly computing the trace we get to
\begin{equation}
\varrho(t + dt ) = \varrho(t) + \kappa \hat c \varrho \hat c^\dag \,dt  - \frac{\kappa}{2} \left\{ \hat c^\dag \hat c , \varrho \right\}  \,dt,
\end{equation}
which we can rewrite as a proper differential equation:
\begin{equation}
\label{eq:Linblad_simple}
\frac{d \varrho(t)}{dt} = \kappa \mathcal{D}[ \hat c] \varrho(t),
\end{equation}
where we introduced the dissipation superoperator $\mathcal{D}$, defined as
\begin{equation}
\label{eq:Dsuperop_def}
\mathcal{D}[\hat A] \bullet = \hat A \bullet \hat A^\dag - \frac{1}{2} \left\{ \hat A^\dag \hat A , \bullet \right\}.
\end{equation}
More in general, we can also insert an Hamiltonian acting on the system into the master equation~\eqref{eq:Linblad_simple}, albeit we should make sure that the evolution due to the system Hamiltonian is slow enough.
In particular, we are making a Markovianity assumption of considering the system ``frozen'' while interacting with each environmental mode: this could break down if the timescale induced by the system Hamiltonian is too short.

Furthermore, the case with a single $\hat c$ can be generalized to a collection $\{\hat c_j\}$, often called collapse operators. 
The general form of a Markovian Lindblad master equation is then
\begin{equation}
\label{eq:Lindblad_generic}
\frac{d \varrho(t)}{dt} = \mathcal{L} \varrho(t) = -i \left[ \hat H , \varrho(t) \right] + \sum_i \kappa_i \mathcal{D}[ \hat c_i] \varrho(t) \,,
\end{equation}
with $\kappa_i \geq 0 \quad \forall i$.
To obtain this general form, we have to consider $n$ independent input modes $\hat b^{(j)}_t$, one for each noise operator $c_j$, that satisfy Bosonic commutation relations 
\begin{equation}
\label{eq:InputCommDeltaMultimode}
\left[ \hat b^{(j)}_t,\hat b^{(k) \dag}_{t'} \right] =\delta_{jk}\delta(t-t') \,.
\end{equation}
In Eq.~\eqref{eq:Lindblad_generic} we also introduced the time independent Lindbladian superoperator $\mathcal{L}$.
This is a \emph{linear} (super)operator and it is easy to see that the complete-positive trace-preserving map (CPTP) generated by Eq.~\eqref{eq:Lindblad_generic} are obtained by exponentiation as as
\begin{equation}
\label{eq:LindbladExp}
\mathcal{E}_t = e^{ t \mathcal{L} } \,.
\end{equation}
These solutions satisfy a semi-group property
\begin{equation}
\label{eq:semigroup_prop}
\mathcal{E}_t \circ \mathcal{E}_s = \mathcal{E}_{s+t} \quad \forall t,s \in \mathbbm{R};
\end{equation}
the fact that this is a semi-group and there is no CPTP inverse map, unlike the purely unitary case, makes clear that this kind of evolution is irreversible.
More in general, if a collection of CPTP maps $\mathcal{E}_t$ satisfies the semi-group property~\eqref{eq:semigroup_prop}, the generator $\mathcal{L}$ must have the form given by the master equation~\eqref{eq:Lindblad_generic}; this is the original result of Lindblad~\cite{Lindblad1976} and Gorini et al.~\cite{Gorini1976}.
\section{Derivation of stochastic master equations}\label{sec:SME}
In the previous section, we have derived the Markovian master equation, assuming that the system interacts with a {\em train} of harmonic oscillators $\{ \hat B_t \}$, that describe the environment, and by tracing out these degrees of freedom after the global interaction.

In the following we will assume that the environment is continuously monitored, that is, after every infinitesimal interaction, the environmental quantum state corresponding to the \emph{quantum oscillator} operator $\hat B_t$ is measured. We will start by focusing on measurements described by projectors on pure states, i.e. by a positive operator-valued measure (POVM) $\{ \Pi_j = | j \rangle \langle j | \} $ and we will later discuss the case of more general (noisy) POVMs.

In the literature, there are several approaches that can be taken in order to derive the corresponding {\em stochastic master equations} (SME) that describes the evolution under time-continuous monitoring. We will exploit what we have already described before for the derivation of the (unconditional) master equation, and we will then follow these steps\footnote{A similar approach can be found in Sec. 4.4.1 of Ref.~\cite{Wiseman1994}, where the SME for continuous homodyne detection in the presence of a generic Gaussian bath is derived.}:
\begin{itemize}
\item 
As before, at each time $t$, the global state of system and environment $R(t)$ is a factorized state with the same reduced state for the environment, i.e.
\begin{align}
R(t) = \varrho^{\text{(c)}}(t) \otimes |0\rangle_t\langle 0| \,, \label{eq:Born}
\end{align} 
where the superscript $(c)$ signals a conditional state.
\item 
The global state evolves in a infinitesimal time $dt$ as
\begin{align}
R(t+dt) = \hat{U}(t,t+dt) R(t) \hat{U}(t,t+dt)^\dag \, \label{eq:evolutionR}
\end{align}
where $\hat{U}(t,t+ dt) = e^{-i \hat{H}_\text{int} (t) \, dt}$.
\item
The conditional state of the system at time $t+dt$ is obtained by projecting the environment on the corresponding projector $|j\rangle\langle j|$ that belongs to the Hilbert space corresponding to the oscillator $\hat{B}_t$, i.e.
\begin{align}
\varrho_j^{\text{(c)}}(t+dt) = \frac{ {}_t\langle j | R(t+dt) | j \rangle_t }{p_j (t+dt) } \,,
\end{align}
where $p_j (t+dt) = \Tr[R(t+dt) (\mathbbm{1}_\mathcal{S} \otimes |j\rangle_t \langle j | )]$ denotes the probability of obtaining the outcome $j$ at time $t+dt$.
\end{itemize}

\subsubsection{Photo-detection stochastic master equation}
We will now consider the case where the environment is continuously monitored with a photo-detector, i.e. by projecting it on the corresponding Fock basis $\{ |n\rangle \langle n|\}$. As it is apparent from Eq.~\eqref{eq:globalev}, one gets non zero probability of detection only by projecting on  the vacuum state $|0\rangle\langle 0|$ or on the single photon state $|1\rangle\langle 1|$. We will thus consider these two results separately:
\begin{itemize}
\item {\bf Measurement result: $\mathbf{0}$}.\\
The (unnormalized) conditional state reads:
\begin{equation}
   \begin{split}
\tilde{\varrho}_0^{\text{(c)}} (t+dt) &= {}_t\langle 0 | R(t+dt) |0 \rangle_t \\
&= \varrho^{\text{(c)}}(t) - \frac{\kappa}2 \{ \varrho^{\text{(c)}}(t) , \hat c^\dag \hat c\} \,dt \,.
   \end{split}
\end{equation}
In order to normalize it, we need to compute the probability:
\begin{align}
p_0(t+dt) = \Tr[ \tilde{\varrho}_0^{\text{(c)}} (t+dt) ] = 1 - \kappa \langle \hat c^\dag \hat c\rangle_t \, dt \,,
\label{eq:p0}
\end{align}
where we have introduced the notation $\langle \hat A \rangle_t = \Tr[\varrho (t) \hat A ]$. By expanding the inverse probability at first order in $dt$, the normalized conditional state reads
\begin{align}
\varrho_0^{\text{(c)}} (t+dt) &= \frac{\tilde{\varrho}_0^{\text{(c)}} (t+dt)}{p_0(t+dt)} \nonumber \\
&\approx \left( \varrho^{\text{(c)}}(t) - \frac{\kappa}2 \{ \varrho^{\text{(c)}}(t) , \hat c^\dag \hat c\} \,dt  \right) \left( 1 + \kappa \langle \hat c^\dag \hat c\rangle_t \, dt \right) \nonumber \\
&\approx  \varrho^{\text{(c)}} - \frac{\kappa}{2} \mathcal{H}[\hat c^\dag \hat c ] \varrho^{\text{(c)}} (t) \, dt \,, \label{eq:PDoutcome0}
\end{align}
where we have introduced the superoperator 
\begin{equation}
\label{eq:Hsuperop_def}
\mathcal{H}[\hat A] \bullet = \hat A \bullet +  \bullet \hat A^\dag - \langle \hat A + \hat A^\dag \rangle_t \bullet  \,.
\end{equation}
\item {\bf Measurement result: $\mathbf{1}$}.\\
The (unnormalized) conditional state reads:
\begin{align}
\tilde{\varrho}_1^{\text{(c)}} (t+dt) = {}_t\langle 1 | R(t+dt) |1\rangle_t = \kappa \hat c \varrho^{\text{(c)}}(t) \hat c^\dag \, dt \,.
\end{align}
In this case the probability reads:
\begin{align}
p_1(t+dt) = \Tr[ \tilde{\varrho}_1^{\text{(c)}} (t+dt) ] = \kappa \langle \hat c^\dag \hat c\rangle_t \, dt \,,
\label{eq:p1}
\end{align}
The normalized conditional state thus reads
\begin{align}
\varrho_1^{\text{(c)}} (t+dt) &= \frac{\tilde{\varrho}_1^{\text{(c)}} (t+dt)}{p_1(t+dt)} \nonumber \\
&= \frac{ \hat c \varrho^{\text{(c)}}(t) \hat c^\dag } {\langle \hat c^\dag \hat c \rangle_t } \,.
\end{align}
\end{itemize}
By observing that the outcome of the measurement corresponds to a random variable taking values $\{0,1\}$ with probabilities given by Eqs. (\ref{eq:p0}) and (\ref{eq:p1}), we can introduce a Poisson increment\footnote{
A Poisson increment $dN_t$ with rate $\lambda(t)$ is a stochastic process with two outcomes: $0$ and $1$. 
The probability that $d N_t$ assumes the value $1$ in the time interval $dt$ is $\lambda (t) dt$, and the probability that it takes the value $0$ is $1- \lambda(t) dt$.
Informally, the probability that $dN_t$ is $1$ in any infinitesimal time interval is vanishingly small and thus $dN$ is zero most of the time.
However, sometimes $d N_t$ takes the value $1$, and the value of $N(T) = \int_0^T d N_t$ ``jumps'' by $1$; $N(t)$ is called a Poisson \emph{process}.
A fundamental property of Poisson increments is that $(dN_t)^2 = dN_t$.
}
$dN_t$ with mean
\begin{equation}
\label{eq:PoissonProcGamma}
\begin{split}
\mathbbm{E}\left[dN_t\right] & = 0  \cdot p_0(t+dt)+ 1 \cdot p_1(t+dt)\\ 
& = \kappa \Tr \left[ \varrho^{\text{(c)}}(t) \hat c^\dag \hat c \right] dt \,,
\end{split}
\end{equation}
where we have introduced the notation $\mathbbm{E}[\bullet]$ to indicate the (stochastic) average of the random variable.
Notice that, being $dN_t$ a two-outcome random variable, it is completely described by its mean value.
Therefore one can combine the two possible outcomes into a single equation
\begin{equation}
\begin{split}
d \varrho^{\text{(c)}}(t) =& \varrho^{\text{(c)}}(t+dt) - \varrho^{\text{(c)}} (t)  \\
=&  d N_t \left( \varrho^{\text{(c)}}_1(t+dt) - \varrho^{\text{(c)}} (t) \right) \\
& + (1 - d N_t) \left( \varrho^{\text{(c)}}_0(t+dt) - \varrho^{\text{(c)}} (t) \right) \,.
\end{split}
\end{equation}
This equation can be simplified by noting that the term $dt dN_t $ can be discarded, since it gives contributions of order greater than $dt$ (this can be understood intuitively from the last two equations).
If we also add an Hamiltonian term (with the same caveats of the previous derivation) we get to the more general equation
\begin{multline}
d\varrho^{\text{(c)}} = -i \left[ \hat H, \varrho^{\text{(c)}} \right]\, dt - \frac{\kappa}{2} \mathcal{H}[\hat c^\dag \hat c] \varrho^{\text{(c)}} dt \\
+\left( \frac{\hat c \varrho^{\text{(c)}} \hat c^\dag }{\Tr \left[\hat c \varrho^{\text{(c)}} \hat c^\dag \right]} - \varrho^{\text{(c)}} \right) d N_t. \label{eq:SME_PD_eff1singleC}
\end{multline}
It is important to notice how the obtained SME is nonlinear in the quantum state $\varrho^{\text{(c)}}$, as both the superoperator $\mathcal{H}[\hat c^\dag \hat c]$ and the statistic of the Poisson process $dN_t$ involve the calculation of the average value $\langle \hat c^\dag c \rangle_t = \Tr[\varrho^{\text{(c)}} \hat c^\dag \hat c]$.
This is a consequence of the renormalization needed to obtain a normalized quantum state, after conditioning on the result of a measurement.

The unconditional state is obtained by averaging over the Poisson process: $\varrho(t) = \mathbbm{E}\left[ \varrho^{\text{(c)}} (t)\right]$.
In order to see it explicitly it is useful to introduce the following stochastic calculus rule~\cite{wiseman2010quantum}  
\begin{equation}
\label{eq:PoissonProp}
\mathbbm{E}\left[ d N_t f(\varrho^{\text{(c)}}) \right] = \kappa\,  dt  \, \mathbbm{E}\left[ \Tr \left[ c^\dag c \varrho^{\text{(c)}} \right] f(\varrho^{\text{(c)}}) \right] \,,
\end{equation}
for any function of the density matrix; this is consistent with~\eqref{eq:PoissonProcGamma} when $f=1$.
By making use of~\eqref{eq:PoissonProp} one can average over the Poisson increment and get from the SME~\eqref{eq:SME_PD_eff1singleC} for the conditional state to the Lindblad master equation~\eqref{eq:Linblad_simple} for the unconditional state, i.e.
\begin{align}
\mathbbm{E}[d \varrho^{\text{(c)}}(t)] =  -i \left[ \hat H, \varrho(t) \right] \,dt+ \kappa \mathcal{D}[\hat c] \varrho(t) \,dt \,.
\end{align}

If instead of considering a generic initial state $\varrho_0$, the system starts in a pure state $\ket{\psi_0}$, the stochastic evolution maintains the state pure at all times and can be rewritten as a (non-linear) equation for a conditional state vector $\ket{\psi_c(t)}$:
\begin{multline}
\label{eq:SSE_PD_singleC}
d \ket{\psi^{\text{(c)}}(t)} = \biggl[  -i \hat{H} dt + \frac{\kappa}{2} \left( \left\langle \hat c^\dag \hat c \right\rangle_t  - \hat c^\dag \hat c\right) {dt} \\ 
+ \biggl( \frac{ \hat c}{\sqrt{\left\langle \hat c^\dag \hat c \right\rangle_t }} - 1 \biggr) dN_t \biggr] \ket{\psi^{\text{(c)}}(t)},
\end{multline}
this equation is called a stochastic Schrödinger equation (SSE); for compactness we introduced the expectation value on the conditional state: $\bra{\psi^{\text{(c)}}(t)} \bullet \ket{\psi^{\text{(c)}}(t)} = \langle \bullet \rangle$.
Historically, solutions to SSEs are called \emph{quantum trajectories}; however we also use the term more freely also for solutions of SMEs.

By writing the infinitesimal increment for the projector representing the pure state 
\begin{multline}
d\varrho^{\text{(c)}} = \left( d \ket{\psi^{\text{(c)}}})\bra{\psi^{\text{(c)}}} +  \ket{\psi^{\text{(c)}}}(d\bra{\psi^{\text{(c)}}} \right)  \\ 
+ \left( d\ket{\psi^{\text{(c)}}})(d\bra{\psi^{\text{(c)}}} \right) \,, \label{eq:drho2ndorder}
\end{multline}
one can check the consistence between~\eqref{eq:SSE_PD_singleC} and~\eqref{eq:SME_PD_eff1singleC}, taking into account the property of Poisson increments $(dN_t)^2 = dN_t$.
Note that it is necessary to retain also the second order term in Eq.~\eqref{eq:drho2ndorder}.

The stochastic variable $N_T = \int_0^T dN_t$ represents how many jumps have occurred up to time $T$, i.e. how many photons have been detected.
However, the `measurement record' corresponding to an experiment contains more information than just $N_T$, the complete information is a list of times $0 \leq t_1, \leq  \dots \leq t_{N_T} \leq T$ corresponding to each photo-detection.

Every master equation in Lindblad form can be rewritten in terms of SSEs and the solution of the master equation is obtained by averaging over the stochastic process.
This procedure is called ``unravelling'' the master equation and Eq.~\eqref{eq:SSE_PD_singleC} is a specific unravelling based on a discontinuous jump-like evolution.
As we have seen, a SSE has a physical meaning, however it is also useful as a tool to numerically solve Lindblad master equations~\cite{Dalibard1992,Plenio1998}.
The advantage is that only state vectors are needed instead of density matrices, the price to pay is that one has to accumulate enough statistics to reduce fluctuations by generating many trajectories.
\subsubsection{Homodyne detection stochastic master equation}
We will now consider the case where the environment is continuously monitored via homodyne detection, i.e. by projecting the environment at each time on the (unnormalized) eigenstates $\{ |x_\theta \rangle \langle x_\theta |\}$ of the operator $\hat x_\theta = (e^{i\theta}\hat B_t + e^{-i \theta} \hat B_t^\dag )/\sqrt{2}$.
In this case the unnormalized conditional state reads
\begin{align}
\tilde{\varrho}^{\text{(c)}}& (t+dt) = \\ 
& = \Tr_t \biggl[ \hat U(t,t+dt) \left( \varrho(t) \otimes \ket{0}_t\bra{0} \right) \hat U(t,t+dt)^\dag \nonumber \\
& \qquad \qquad \qquad \qquad \qquad \left( \id \otimes \ket{x_\theta}_t \! \bra{x_\theta} \right) \biggr] = \nonumber \\
&= {}_t\bra{x_\theta} \hat U(t,t+dt) \left( \varrho(t) \otimes \ket{0}_t \! \bra{0} \right) \hat U(t,t+dt)^\dag \ket{x_\theta}_t \nonumber
\end{align}
and the corresponding probability
\begin{equation}
p_{x_\theta} (t+dt) = \Tr \left[ \tilde{\varrho}(t+dt) \right] \,;
\end{equation}
the normalized state is then
\begin{equation}
\label{eq:norm_unnorm_condstate}
\varrho^{\text{(c)}}(t+dt)= \frac{\tilde{\varrho}^{\text{(c)}}(t+dt)}{p_{x_\theta}(t+dt)} \,.
\end{equation}
Differently from the previous case, we only need an expansion of the numerator and the denominator in~\eqref{eq:norm_unnorm_condstate} up to order $\sqrt{\kappa dt}$, since all the relevant terms of order $\kappa dt$ come out as a result.
Due to vacuum fluctuations the probability distribution for $x_\theta$ at time $t$, before the interaction, is a Gaussian with variance $\frac{1}{2}$ and mean value $0$, i.e.  $p_{x_\theta}(t) = \left\vert {}_t\langle x_\theta \vert 0 \rangle_t \right\vert^2 = \frac{1}{\sqrt{\pi}} e^{-x_\theta^2}$.

The expansion of the conditional state yields
\begin{align}
   \label{eq:rhocondHomo}
\tilde{\varrho}^{\text{(c)}}\!(t+dt) &=  \left\vert {}_t\langle x_\theta \vert 0 \rangle_t \right\vert^2 \varrho^{\text{(c)}}(t) \\
& \quad + \biggl( \hat c \varrho^{\text{(c)}}(t) {}_t\langle x_\theta | 1 \rangle_t \langle 0 \vert x_\theta \rangle_t + \nonumber \\ 
& \qquad \varrho^{\text{(c)}}(t) \hat c^\dag {}_t\langle x_\theta \vert 0 \rangle_t \langle 1 \vert x_\theta \rangle_t \biggr) \sqrt{\kappa dt} + o(\sqrt{ \kappa dt}) \nonumber \\
=  p_{x_\theta}(t) & \biggl[  \varrho^{\text{(c)}}(t) \nonumber \\ 
& \mkern-36mu  +   \left( \hat c \varrho^{\text{(c)}}(t) e^{i \theta} + \varrho^{\text{(c)}}(t) \hat c^\dag  e^{-i \theta}  \right) x_\theta \sqrt{2 \kappa dt} \biggr] + o(\sqrt{\kappa dt}) \nonumber
\end{align}
where to get the second expression we used the identity ${}_t\langle x_\theta | 1  \rangle_t = \sqrt{2}x_\theta e^{i\theta} {}_t\langle x_\theta | 0  \rangle_t$.
The expansion of the probability distribution yields
\begin{align}
& p_{x_\theta}(t+dt) = \Tr\left[ \tilde{\varrho}^{\text{(c)}}(t+dt) \right] \label{eq:pdt1} \\
 & \qquad = p_{x_\theta}(t) \left(1 + \sqrt{2 \kappa dt} x_\theta \langle \hat c e^{i \theta} + \hat c^\dag e^{-i \theta} \rangle_t \right) + o(\sqrt{\kappa dt}) \nonumber \\
& \qquad \approx  \frac{1}{\sqrt{\pi}} \exp \left[-\left(x_\theta - \sqrt{\frac{\kappa dt}{2}} \left\langle \hat c e^{i \theta} + \hat c^\dag e^{-i \theta} \right\rangle_t \right)^2 \right] , \nonumber
\end{align}
up to order $\sqrt{dt}$ it is equivalent to a Gaussian distribution with variance $\frac{1}{2}$ and mean value $\sqrt{\frac{\kappa dt}{2}} \langle \hat c e^{i \theta} + \hat c^\dag e^{-i \theta} \rangle_t$.
We can introduce a new stochastic increment $d y_t$ and a stochastic process $I(t)$ such that
\begin{equation}
\label{eq:dySME}
d y_t = I(t) dt = \sqrt{2 dt} x_\theta = \sqrt{\kappa} \langle \hat c e^{i \theta} + \hat c^\dag e^{-i \theta} \rangle_t dt + dw_t \,,
\end{equation}
where here $dw_t$ a Gaussian random variable with variance $dt$ and zero mean $\mathbbm{E}\left[ d w_t \right] = 0$.
The stochastic variable $y_T = \int_0^T d y_t$ represents the so-called integrated photo-current.
The continuous stream of outcomes is usually understood in terms of the so-called photocurrent $I(t) = \frac{d y_t}{dt}$.
The photocurrent $I_t$ is equal to the expectation value of the operator $\hat c e^{i \theta} + \hat c^\dag e^{-i \theta}$ at time $t$, plus a white noise term:
\begin{equation}
I(t) = \sqrt{\kappa} \langle \hat c e^{i \theta} + \hat c^\dag e^{-i \theta} \rangle_t + \xi(t)
\end{equation}
where $\xi(t) = \frac{dw_t}{dt}$ represents white-noise, that is typically mathematically (heuristically) defined by the condition $\mathbbm{E}[\xi(t) \, \xi(t')]=\delta(t-t')$.

In the theory of stochastic processes $dw_t$ is called a Wiener increment.
The defining and remarkable property of a Wiener increment is that its square (in principle a random variable) is actually \emph{not} random, but satisfies $dw_t^2 = dt$ deterministically\footnote{A proof of this identity can be found in Sec. 5.2 of Ref.~\cite{Jacobs2006} or in Appendix C of Ref.~\cite{Serafini2023}.}.
This identity is formally known as Itô's lemma and it is the basis of Itô stochastic calculus\footnote{In this work we only use Itô calculus and the currents $I(t)$ and $\xi(t)$ that we have formally written as time derivatives are, strictly speaking, not well-defined and should should be implicitly understood through the increment definitions, e.g. $ \xi(t) dt = d w_t$.
Note, however, that it is alternatively possible to employ stochastic differential equations in Stratonovich form, in which the chain rule of usual calculus holds and currents such as $\xi(t)$ are well-defined objects. 
See Refs.~\cite{wiseman2010quantum,Jacobs2014a} for a discussion of this difference and its physical interpretation.}.
The assumption leading to a Itô stochastic differential equation is that integral~\eqref{eq:int_bin}, needed to define the ``physical'' operators $\hat b'_\text{in}(t)$, does not involve modes at times preceding $t$.

By inserting expressions~\eqref{eq:rhocondHomo} and~\eqref{eq:pdt1} into~\eqref{eq:norm_unnorm_condstate} and expanding the denominator we can finally get to
\begin{align}
d \varrho^{\text{(c)}} (t) &=  \sqrt{\kappa} \mathcal{H}[ \hat c e^{i\theta}] \varrho^{\text{(c)}}(t) \, dw_t + o(\sqrt{\kappa dt}) \,.
\end{align}
Since averaging over the measurement result, i.e. over the stochastic Wiener increment $dw_t$ should yield the unconditional Markovian master equation \eqref{eq:Linblad_simple}, the remaining terms of order $O(dt)$ are obtained straightforwardly, and we obtain the SME:
\begin{align}
\label{eq:SMEHomoSymple}
d \varrho^{\text{(c)}} (t) =& -i \left[ \hat H, \varrho^{\text{(c)}}(t) \right] dt + \kappa \mathcal{D}\left[ \hat c \right] \varrho^{\text{(c)}}(t) dt \\
& \qquad \qquad \qquad \qquad +  \sqrt{\kappa} \mathcal{H}[ \hat c e^{i\theta}] \varrho^{\text{(c)}}(t) \,dw_t  \nonumber \\
=& -i\left[ \hat H, \varrho^{\text{(c)}}(t) \right] dt + \kappa \mathcal{D}\left[ \hat c \right] \varrho^{\text{(c)}}(t) dt  \label{eq:SMEHomoSymple_dy} \\
& + \sqrt{\kappa} \mathcal{H}[ \hat c e^{i\theta}] \varrho^{\text{(c)}}(t) \left( dy_t - \sqrt{\kappa} \langle \hat c e^{i\theta}  + \hat c^\dag e^{-i\theta}\rangle_t dt \right) \nonumber
\end{align}
where we also added the usual Hamiltonian term and in the second line we wrote the equation in term of the observed photocurrent.
As in the photo-detection case, also this SME is evidently non-linear in the quantum state $\varrho^{\text{(c)}}$, involving the calculation of the average value $\langle \hat c e^{i\theta} + \hat c^\dag e^{-i\theta}\rangle_t = \Tr[\varrho^{\text{(c)}} (\hat c e^{i\theta} + \hat c^\dag e^{-i\theta})]$, as a consequence of the renormalization of the conditional state due to the measurement.
This SME is called a diffusive unravelling, since the stochastic part is represented by a diffusive process\footnote{\emph{Diffusive} means that something initially localized spreads out in time. This can be seen by considering the Wiener process $W(t) \equiv \int d w_t $, which is a Gaussian random variable with variance $t$ (linearly increasing in time).}.

Beyond homodyne detection, we can also obtain the SME for continuous heterodyne detection, which mathematically corresponds to projections on coherent states of the operators $\{\hat{B}_t\}$, simply by considering an unravelling with two distinct jump operators $\hat{c}_1=\hat{c}/\sqrt{2}$ and $\hat{c}_2 = i \hat{c}/\sqrt{2}$, corresponding to homodyne phases equal to $\theta=0$ and $\theta=\pi/2$ respectively, yielding (for $\hat{H}=0$)
\begin{align}
d \varrho^{\text{(c)}} (t) =& \frac{\kappa}2 \mathcal{D}\left[ \hat c \right] \varrho^{\text{(c)}}(t) dt  +  \sqrt{\frac{\kappa}2} \mathcal{H}[ \hat c ] \varrho^{\text{(c)}}(t) \,dw^{(1)}_t   \\
& + \frac{\kappa}2 \mathcal{D}\left[ i \hat c \right] \varrho^{\text{(c)}}(t) dt  +  \sqrt{\frac{\kappa}2} \mathcal{H}[ i \hat c ] \varrho^{\text{(c)}}(t) \,dw^{(2)}_t \nonumber \\
=& \kappa \mathcal{D}\left[ \hat c \right] \varrho^{\text{(c)}}(t) dt  +  \sqrt{\frac{\kappa}2} \mathcal{H}[ \hat c ] \varrho^{\text{(c)}}(t) \,dw^{(1)}_t  \nonumber \\
& \qquad \qquad \qquad +  \sqrt{\frac{\kappa}2} \mathcal{H}[ i \hat c ] \varrho^{\text{(c)}}(t) \,dw^{(2)}_t  \,,  \nonumber
\end{align} 
where in the last line we have used the general property $\mathcal{D} \left[ \hat{c} \right] = \mathcal{D} \left[ \hat{c} e^{i \theta} \right]$ and where $dw_t^{(1)}$ and $dw_t^{(2)}$ are uncorrelated Wiener increments related to the two photocurrents:
\begin{equation}
\begin{split}
d y_t^{(1)} & = \sqrt{\frac{\kappa}{2}} \langle \hat c + \hat c^\dag \rangle_t dt + dw_t^{(1)} \, , \\
d y_t^{(2)} & = \sqrt{\frac{\kappa}{2}} \langle i (\hat c - \hat c^\dag )\rangle_t dt + dw_t^{(2)} \, .  
\end{split}
\end{equation}
This formulation is easy to understand reminding ourselves how heterodyne detection can be indeed implemented as a double homodyne detection~\cite{Genoni2014a}.

In the following we will focus on homodyne detection only, and suppress the explicit dependence on $\theta$ which comes from measuring a different quadrature, since it is simply equivalent to the substitution $\hat c \mapsto \hat c e^{i \theta} $, and, as we noted above, the dissipative part does not depend on $\theta$.
Similarly to the discontinuous jump-like case we can alternatively write a SSE if the initial state is pure:
\begin{multline}
d \ket{\psi^{\text{(c)}}(t)} = \biggl\{ \Bigl[ - i \hat{H}  - \frac{\kappa}{2}\bigl( \hat{c}^\dag \hat{c} - \langle \hat c + \hat{c}^\dag \rangle \hat{c}  + \frac{\langle \hat c + \hat c^\dag \rangle^2}{4} \bigr) \Bigr] dt  \\ 
+ \sqrt{\kappa} \bigl( \hat c -  \frac{\left\langle \hat{c} + \hat{c}^\dag \right\rangle}{2} \bigr) dw_t \biggr\} \ket{\psi^{\text{(c)}}(t)} \, ;
\end{multline}
to check the equivalence with the previous equation for the density matrix we need to use again Eq.~\eqref{eq:drho2ndorder} and to apply Itô's rule.
Due to the fact that $d w_t$ has zero mean $\mathbbm{E}\left[ d w_t \right]=0$, it is straight forward to check that the average of~\eqref{eq:SMEHomoSymple} gives back the Lindblad equation~\eqref{eq:Linblad_simple} (written there for the particular case $\hat H = 0$). 

\subsubsection{Inefficient detection}
The previous equations were derived assuming that the measurement on the output modes was a perfect projective measurement (actually a sharp observable, since we are only interested in the statistics).
However this assumption is not fulfilled in many practical and interesting scenarios.
The simplest way to model inefficient detection is to assume that the measurement happens with probability $\eta \in [0,1]$ and fails with probability $1-\eta$.
This is equivalent to placing a beam-splitter with transmissivity $\eta$ before the measurement device, which accounts for the lost photons.
Another equivalent way to model this inefficiency is to rewrite the master equation for the unconditional dynamics as
\begin{equation}
\frac{d \varrho (t)}{dt }= \eta \kappa \mathcal{D}\left[ \hat c \right] \varrho(t) + (1-\eta) \kappa \mathcal{D}\left[ \hat c \right] \varrho(t)
\end{equation}
and then unravel only the first term proportional to $\eta$~\cite{wiseman2010quantum}.

The steps presented in the previous sections can be reproduced and we get to a SME for photo-detection
\begin{multline}
   d \varrho^{\text{(c)}} = -i \left[ \hat H, \varrho^{\text{(c)}} \right] dt -  \frac{\eta\kappa}{2} \mathcal{H}[\hat c^\dag \hat c] \varrho^{\text{(c)}} dt \label{eq:SME_PD_singleC}  \\
 + \biggl( \frac{ \hat c \varrho^{\text{(c)}} \hat c^\dag }{\Tr \left[ \hat c \varrho^{\text{(c)}} \hat c^\dag \right]} - \varrho^{\text{(c)}} \biggr) d N_t + (1-\eta) \kappa \mathcal{D}[\hat c]\varrho^{\text{(c)}} dt \,,
\end{multline}
where now the Poisson increment has expectation value $\mathbbm{E}\left[ dN_t \right] = \eta \kappa \Tr\left[ \varrho^{\text{(c)}} \right] dt$.
Analogously, we have a diffusive SME for homodyne detection with finite efficiency
\begin{multline}
d \varrho^{\text{(c)}} (t) =  -i \left[ \hat H, \varrho^{\text{(c)}}(t) \right] dt +  \kappa \mathcal{D}\left[ \hat c \right] \varrho^{\text{(c)}}(t) dt  \\
 + \sqrt{\eta\kappa} \mathcal{H}[ \hat c e^{i\theta}] \varrho^{\text{(c)}}(t) dw_t,
\end{multline}
corresponding to a measured continuous photo-current $dy_t = \sqrt{\eta\kappa} \langle \hat{c} + \hat c^\dag \rangle_t \,dt + dw_t$.
As expected, both equations give back the unconditional Lindblad master equation for $\eta \to 0$, as it physically corresponds to not performing any measurement on the output operators.
\subsubsection{Linear quantum trajectories}
\label{subsec:linQtraj}
The evolution of a quantum system conditioned on a certain measurement outcome is given by a trace non-increasing linear CP map.
On the other hand, as observed above, the map which gives the \emph{normalized} conditional state is non-linear, because of the renormalization. 
As we will see in the following, it is possible to derive \emph{linear} SMEs describing the evolution of the unnormalized conditional state, that have the advantage of being easier to solve analytically and numerically, and thus to obtain an explicit time evolution operator corresponding to the measurement results $dN_t$ and $dy_t$.
We will start here by focusing on the homodyne detection SME for unit detection $\eta=1$ (everything can be readily extended to all values of $\eta$), where the nonlinearity is due to the scalar coefficient $\Tr \left[ (\hat{c}+ \hat{c}^\dag ) \varrho^{\text{(c)}} \right]$.
By applying the formal substitution $\Tr\left[ \left( \hat{c}  + \hat c^\dag  \right)\varrho^{\text{(c)}} \right] \to \mu$, one obtains a linear SME for the (unnormalized) density operator $\bar{\varrho}$:
\begin{multline}
\label{eq:linearSMEhomoSingleC}
d\bar{\varrho}^{\text{(c)}} =  \left( -i \left[ \hat H, \bar{\varrho}^{\text{(c)}}\right] + \kappa\mathcal{D}\left[c  \right] \bar{\varrho}^{\text{(c)}} \right) dt  \\
 + \sqrt{\kappa}\left( \hat c \bar{\varrho}^{\text{(c)}} + \bar{\varrho}^{\text{(c)}} \hat c^\dag - \mu \bar{\varrho}^{\text{(c)}} \right) \left( d y_t -  \sqrt{\kappa} \mu \,dt \right), 
\end{multline}
that, for the typical choice $\mu=0$ reads
\begin{multline}
d\bar{\varrho}^{\text{(c)}} = \left( -i \left[ \hat H, \bar{\varrho}^{\text{(c)}}\right] + \kappa \mathcal{D}\left[c  \right] \bar{\varrho}^{\text{(c)}} \right) dt \\
 + \sqrt{\kappa} \left( \hat c \bar{\varrho}^{\text{(c)}} + \bar{\varrho}^{\text{(c)}} \hat c^\dag \right) d y_t \,,
\label{eq:linearSMEmu0}
\end{multline}
Two observations are now in order. 
\begin{itemize}
\item
By calculating the trace of the evolved operator $\bar{\varrho}^{\text{(c)}}(t+dt) = \bar{\varrho}^{\text{(c)}}(t) + d\bar{\varrho}^{\text{(c)}}(t)$, one can show that the renormalized conditional operator can be calculated at each time $t+dt$ as
\begin{align}
\varrho^{\text{(c)}}(t+dt) = \frac{\bar{\varrho}^{\text{(c)}}(t+dt)}{\Tr[\bar{\varrho}^{\text{(c)}}(t+dt)]} \,.  \label{eq:barrho}
\end{align}
\item
The unconditional Markovian master equation (\ref{eq:Linblad_simple}) can be obtained from the linear SME above, if the random variable $dy_t$ is chosen according to a, so-called, \emph{ostensible} probability\footnote{The term ``ostensible probability'' has been introduced by Wiseman~\cite{Wiseman1996,wiseman2010quantum}; it may be clearer to call it a reference probability, as in Ref.~\cite{Gammelmark2013a}.}, that does not depend on the conditional state, and that in this case corresponds to a Gaussian random variable, centered in $\sqrt{\kappa} \mu \,dt$ and with variance $dt$. In formula (for $\hat{H}=0$): 
\begin{align}
\mathbbm{E}_{p_\text{ost}} [ d\bar{\varrho}^{\text{(c)}}(t)] =  \kappa \mathcal{D}[\hat c] \varrho(t) \, dt \,.
\end{align}
In the formula above we have explicitly written that the stochastic average has to be taken according to the ostensive probability. 
\end{itemize}
Consequently, by denoting with $p_\text{true}(dy_t)$ the true probability distribution of $dy_t$ (that is, a Gaussian distribution centered in $\sqrt{\kappa}\langle \hat c + \hat c^\dag \rangle_t \,dt$ and with variance $dt$), we can write the unconditional density operator, solution of (\ref{eq:Linblad_simple}):
\begin{align}
\varrho_\text{unc}(t) &= \mathbbm{E}_{p_\text{true}} [ \varrho^{\text{(c)}}(t)] = \int dJ \, p_\text{true} (J) \,\varrho^{\text{(c)}}(t)  \,, \\
\varrho_\text{unc}(t) &= \mathbbm{E}_{p_\text{ost}} [ \bar{\varrho}^{\text{(c)}}(t) ] = \int dJ \, p_\text{ost} (J)  \, \bar{\varrho}^{\text{(c)}}(t) \,,
\end{align}
where we have explicitly written the stochastic average and, for typographic reasons, we denoted the measurement outcome as $dy_t=J$. These equations, together with  Eq. (\ref{eq:barrho}), yield the relationship between the true and the ostensive probability:
\begin{align}
p_\text{true}(dy_t) = p_\text{ost}(dy_t) \Tr[\bar{\varrho}^{\text{(c)}}(t)] \,, 
\label{eq:truevsostensive}
\end{align}
showing how the trace of the solution of the linear quantum trajectory $\Tr[\bar{\varrho}^{\text{(c)}}(t)]$ encodes the probability of the conditional state $\varrho^{\text{(c)}}(t)$.

An analogous discussion can be made for the SMEs~\eqref{eq:SME_PD_eff1singleC} for photo-detection, where the nonlinear terms are due to the scalar coefficient $\Tr [ \hat{c}^\dag \hat{c} \varrho^{\text{(c)}} ]$.
In this case, if we perform the substitution $\Tr [ \hat{c}^\dag \hat{c} \varrho^{\text{(c)}} ] \to \beta$ for any $\beta > 0$ we get the following linear equation
\begin{multline}
\label{eq:linSMEpd}
d\bar{\varrho}^{\text{(c)}} =  \left( -i \left[ \hat{H}, \bar{\varrho}^{\text{(c)}}\right] - \frac{\kappa}{2} \left\{ \hat{c}^\dag \hat{c} ,  \bar{\varrho}^{\text{(c)}} \right\} + \beta \kappa \bar{\varrho}^{\text{(c)}}  \right) dt  \\
 + \left( \frac{\hat{c} \bar{\varrho}^{\text{(c)}} \hat{c}^\dag}{\beta}  - \bar{\varrho}^{\text{(c)}} \right) dN_t,
\end{multline}
that, for the usual choice $\beta = 1$, reads,
\begin{multline}
d\bar{\varrho}^{\text{(c)}} =  \left( -i \left[ \hat{H}, \bar{\varrho}^{\text{(c)}}\right] - \frac{\kappa}{2} \left\{ \hat{c}^\dag \hat{c} ,  \bar{\varrho}^{\text{(c)}} \right\} + \kappa \bar{\varrho}^{\text{(c)}}  \right) dt \\
 + \left( \hat{c} \bar{\varrho}^{\text{(c)}} \hat{c}^\dag  - \bar{\varrho}^{\text{(c)}} \right) dN_t,
\end{multline}
As before, one can prove that the (renormalized) conditional state, solution of the non-linear SME~\eqref{eq:SME_PD_eff1singleC}, can be obtained via the solution of the linear SME~\eqref{eq:linSMEpd}, via the formula $\varrho^{\text{(c)}}(t) = \bar{\varrho}^{\text{(c)}}/\Tr[\bar{\varrho}^{\text{(c)}}]$; moreover one observes that the unconditional Markovian master equation in this case is obtained by averaging the linear SME~\eqref{eq:linSMEpd} over an ostensible probability, such that $\mathbbm{E}_{p_\text{ost}}[dN_t] =\eta \kappa \beta \,dt$.
As a consequence, one still obtain the relationship $p_\text{true}(dN_t) = p_\text{ost}(dN_t) \Tr[\bar{\varrho}^{\text{(c)}}(t)]$ between true and ostensible probability distributions.
\subsubsection{Completely positive infinitesimal evolution - description in terms of (generalized) Kraus operators}
The most natural approach for solving SMEs is to write down a system of coupled stochastic differential equations for the matrix elements of the density operator and then use existing numerical methods.
It might also be useful to consider the Bloch form of the equation, obtained by expanding the density matrix on a basis of Hermitian operators, instead of the matrix elements in the canonical basis, so that all the coefficients are real.
Unfortunately, due to the stochastic nature of the problem there is no guarantee that the evolved state is always perfectly positive, even when using rather advanced numerical methods.
For these reasons it is very useful to employ a numerical method that completely preserves the positivity of the conditional state.
In this way it is also possible to get sensible results without the need of advanced numerical solvers.
The approach that we present was introduced by Rouchon~\cite{Rouchon2014,Rouchon2015} with the goal to achieve stability and speed, in order to enable \emph{real-time} tracking of monitored systems during actual experiments.

The main insight of this method is to notice that the evolution of the conditional state in an infinitesimal time step $dt$ can always be written as the action of a CP map.
This point of view comes naturally from the idea that we can model each infinitesimal evolution in terms of Kraus operators. We will mainly focus on the case of a diffusive evolution due to continuous homodyne monitoring.
In this case the evolution corresponding to the SME \eqref{eq:SMEHomoSymple_dy} can be obtained via the formula
\begin{align}\label{eq:rouchon}
\varrho^{\text{(c)}}(t+dt) = \frac{ \hat M_{dy_t} \varrho^{\text{(c)}}(t) \hat M_{dy_t}^\dag + \kappa (1-\eta) \hat{c} \varrho^{\text{(c)}}(t) \hat{c}^\dag \,dt} { \Tr[ \hat M_{dy_t}\varrho^{\text{(c)}}(t) \hat M_{dy_t}^\dag +  \kappa (1-\eta)  \hat{c} \varrho^{\text{(c)}}(t) \hat{c}^\dag \,dt ]} \, ,
\end{align}
where the Kraus operator reads
\begin{align}
\hat M_{dy_t}= \mathbbm{1} - i \hat{H} \,dt - \frac{\kappa}{2}  \hat{c}^\dag \hat{c} \, dt +  \sqrt{\eta\kappa} \hat{c} \,dy_t \,, \label{eq:Mdyvec}
\end{align}
with $dy_t = \sqrt{\eta \kappa}\,\Tr[\varrho_t^{\text{(c)}} (\hat{c} + \hat{c}^\dag)] \,dt + dw_t$ being the measured infinitesimal photocurrent. 

We remark that the numerator in Eq.~\eqref{eq:Mdyvec} corresponds to the state $\bar{\varrho}_c(t+dt)$ evolved according to the linear SME in Eq.~\eqref{eq:linearSMEmu0}.
As a consequence its trace does not correspond to the true probability, which can be obtained by multiplying it to the corresponding \emph{ostensible} probability as in Eq.~\eqref{eq:truevsostensive}.
For this reason, the operators $\{\hat{M}_{dy_t}\}$ do not correspond to a set of \emph{normalized} Kraus operators, i.e. the corresponding map is not trace preserving, since
\begin{align}
\int dJ\, (\hat{M}_{J}^\dag \hat{M}_{J}) \neq \hat{\mathbbm{1}}  + O(dt^2)  \,,
\end{align}
To obtain the correct normalization one has indeed to include the ostensive probability in the integral, yielding
\begin{align}
\int dJ\, p_{\sf ost}(J) \,(\hat{M}_{J}^\dag \hat{M}_{J}) = \hat{\mathbbm{1}} + O(dt^2) \,,
\end{align}
where, as in the previous section for typographic reasons we have denoted the measurement outcome as $dy_t=J$, and where in this case we remind that $p_{\sf ost}(J)$ is simply a Gaussian centered in zero and with variance $dt$.
Despite the above observation, in the following we will keep referring to the operators $\{\hat{M}_{dy_t}\}$ as Kraus operators with this wider meaning, as the operatorial form of the numerator is sufficient to ensure the complete positivity of the map, and thus the positivity of the evolved state $\varrho^{\text{(c)}}(t+dt)$.

Equation~\eqref{eq:rouchon} can be used for numerical purposes by replacing the infinitesimal increment $dt$ with a finite time step $\Delta t$, the Wiener increments $dw_t$ are then replaced by Gaussian random variables $\Delta w$ centered in zero and with variance equal to $\Delta t$ (one should notice, that for finite increment $\Delta t$, the deterministic identity does not hold, i.e. $\Delta w^2 \neq \Delta t$, this being the main issue with the standard numerical approaches used to solve the homodyne SMEs).
It is also possible to introduce higher-order corrections, either in the spirit of the Euler-Milstein method as in Ref.~\cite{Rouchon2014,Rouchon2015}, or following the approach of Ref.~\cite{Guevara2018} that guarantees complete positivity of the conditional dynamics to order $(\Delta t)^2$.

One can appreciate that in the case of unit efficiency $\eta = 1$ the evolution is given by a single rank one Kraus operator, but when we have finite efficiency the map applied to $\varrho^{\text{(c)}}(t)$ is a CP map.
As we have anticipated before, the numerator of~\eqref{eq:rouchon} is in Kraus form, i.e. it is a sum of operators acting on the left with their conjugates acting on the right.
Therefore, even for finite $\Delta t$ the evolved state at $t+\Delta t$ is always positive, and we are reassured that density operators are sent into density operators.

For photodetection one only need the two Kraus operators
\begin{align}
\label{eq:M0pd} \hat M_0 &= \mathbbm{1} - i \hat{H} \, dt - \frac{\kappa}{2} \hat{c}^{\dag} \hat{c}^{} \,dt \, ,\\
\label{eq:M1pd} \hat M_{1} &= \sqrt{\eta\kappa } \hat{c} \, \sqrt{dt} \,,
\end{align}
where $M_0$ corresponds to the ``no detector click'' event, and the operator $\hat M_1$ to the detection of one photon.
At each time step, each $\hat M_{1}$ has to be applied with probability $p_1 = \eta \kappa \Tr[\varrho^{\text{(c)}}(t) \hat{c}^\dag \hat{c} ]\,dt$ and, correspondingly, the Kraus operator $\hat M_0$ has to be applied with probability $p_0 = 1 - p_1$.
Even in this case the numerical algorithm is obtained by simply replacing $dt$ with a finite $\Delta t$ and the evolution applied at each step is a CP map preserving the positivity of the state.
\section{Markovian feedback master equations}\label{sec:feedback}
\label{s:MarkovianFME}
In the previous section, we have obtained the SMEs describing the evolution of a quantum state that is continuously monitored via continuous {\em weak} measurement on the system, implemented by measuring a Markovian environment interacting with the system.
We have shown how, keeping track of these measurement results (typically dubbed photo-currents), one can keep track of the corresponding conditional quantum state. For an initial pure state and in the case of perfect measurement efficiency, the quantum state stays pure during the whole evolution.
However, in many experimental implementations, it would be desirable to exploit the information obtained from the continuous monitoring in order to engineer, typically at steady-state, an unconditional deterministic quantum state.
This is done by acting with some feedback operation directly on the quantum system.
The kind of feedback that one can apply is typically divided in two main categories: {\em state-based feedback} and {\em Markovian feedback}.

State-based feedback represents the most general kind of feedback that one can think of: by denoting with $I(t)$ the photocurrent obtained via the time-continuous measurement (e.g. $I(t)=dN_t/dt$ for photo-detection, and $I(t) = dy_t / dt$ for homodyne detection), state-based feedback exploits the whole history of the measurement results, i.e.  the photocurrents $\{ I(t')\}_{t'=0}^{t}$ during the whole experiment from time $t'=0$ to $t'=t$, in order to decide the feedback operation to be applied.
State-based feedback is sometimes referred to as {\em Bayesian feedback}; the term {\em Bayesian} comes from the fact that, thanks to this information, one decides the optimal feedback strategy after having updated in a {\em Bayesian way} the conditional state of the quantum system. This kind of feedback is evidently very general, but, from an experimental point of view, it may be extremely difficult to implement: it is in fact necessary to run the algorithm to update the conditional state $\varrho^{\text{(c)}}$ on a time scale much faster than the time scale of the evolution of the quantum state itself, and thus of the time-continuous monitoring.

Markovian feedback on the other hand exploits only the last measurement result $I(t)$, that is fed back into a corresponding feedback Hamiltonian. In the following we will only discuss this case, and we will focus on linear feedback Hamiltonians, such that
\begin{align}
\hat{H}_\text{fb} (t) = \hat{F} I(t)
\end{align}
where $\hat F$ is a fixed (pre-determined) feedback operator. In order to obtain the stochastic evolution of the conditional state under the action of the (Markovian) feedback operation $\varrho^{(\fb)}(t)$ from time $t$ to time $t+dt$, we will exploit the description of the SME in terms of Kraus operators (that we will denote here as $\hat{M}_{I(t)}$, showing explicitly their dependence on the measured photocurrent $I(t)$) and we will follow the ({\em maybe trivial, but fundamental}) intuition that the feedback operation has to be applied {\em after} the action of the measurement operator, i.e.
\begin{align}
\varrho^{\text{(c,fb)}} (t + dt) = \frac{ \hat U_\text{fb} \hat M_{I(t)} \, \varrho^{\text{(c,fb)}}(t)  \, \hat M_{I(t)}^\dag \hat U_\text{fb}^\dag } {\Tr\left[ \hat M_{I(t)} \, \varrho^{\text{(c,fb)}}(t)  \, \hat M_{I(t)}^\dag\right]} \,, \label{eq:feedback_dynamics}
\end{align}
where we have introduced the unitary feedback operation $\hat U_\text{fb} = \exp \{ -i \hat H_\text{fb} \,dt\}$.

We will now derive explicitly the equations describing these evolutions for the case of continuous photo-detection and homodyne detection. As the goal of these feedback protocols is typically to prepare {\em unconditionally} a deterministic target state (or with a given target property), we will also derive the Markovian feedback master equations that describe the evolution obtained by averaging over all the possible trajectories.
We remark that Eq.~\eqref{eq:feedback_dynamics} assumes the implementation of an instantaneous feedback operation; we refer to Ref.~\cite{wiseman2010quantum} for the derivation of feedback master equations for a finite time-delay in the feedback implementation.
\subsubsection{Markovian feedback for continuous photo-detection}
In this case the (Markovian) feedback Hamiltonian is $\hat{H}_\text{fb} (t) = \hat{F} I(t) = \hat{F} (dN_t / dt )$, and the corresponding unitary feedback operator reads 
$$
\hat U_\text{fb} = e^{-i \hat{H}_\text{fb} (t)\,dt} = e^{-i \hat{F} dN_t},
$$ 
where $dN_t$ is a Poissonian increment taking values $dN_t =0$ or $dN_t=1$. 
As for the derivation of the SME that we have presented in the previous section, we will study separately the effect of feedback on the two possible conditional states.
\begin{itemize}
\item {\bf Measurement result: $\mathbf{0}$}.\\
In this case, as $dN_t=0$, the feedback operator is equal to the identity operator $\hat U_\text{fb} = \mathbbm{1}$.  As a consequence the normalized conditional state is simply equal to
\begin{align}
\varrho_0^{\text{(c,fb)}} (t+dt) =  \varrho^{\text{(c,fb)}}(t) - \frac{\kappa}{2} \mathcal{H}[\hat c^\dag \hat c ] \varrho^{\text{(c,fb)}}(t) \, dt \,,
\end{align}
as in Eq.~\eqref{eq:PDoutcome0}.
\item {\bf Measurement result: $\mathbf{1}$}.\\
If a jump is detected, i.e. for $dN_t = 1$, the feedback unitary operator reads $\hat U_\text{fb}  = e^{- i \hat{F}}$. As a consequence we can derive the unnormalized conditional state, after the feedback operation, as
\begin{align}
\tilde{\varrho}_1^{\text{(c,fb)}} (t+dt) &= \hat U_\text{fb} \hat{M}_1 \varrho^\text{(c,\fb)}(t) \hat{M}_1^\dag \hat U_\text{fb}^\dag \nonumber \\
&= \hat U_\text{fb} \left( \hat{c} \varrho^\text{(c,fb)}(t) \hat{c}^\dag \right) \hat U_\text{fb}^\dag \nonumber \\
&= \left( e^{-i \hat{F}} \hat{c} \right) \varrho^\text{(c,fb)} \left(e^{-i \hat{F}} \hat{c} \right)^\dag .
\end{align}
By exploiting the unitarity of the feedback operator one finds that the probability does not change, i.e.
\begin{align}
p_1(t+dt) = \Tr[ \tilde{\varrho}_1^{\text{(c,fb)}} (t+dt) ] = \kappa \langle \hat c^\dag \hat c\rangle_t \, dt \,,
\label{eq:p1_feedback}
\end{align}
The normalized conditional state thus reads
\begin{align}
\varrho_1^{\text{(c,fb)}}(t+dt) &= \frac{\tilde{\varrho}_1^{\text{(c,fb)}} (t+dt)}{p_1(t+dt)} \nonumber \\
&= \frac{ ( e^{-i \hat{F}} \hat{c} ) \varrho^\text{(c,fb)}(t) ( e^{-i \hat{F}} \hat{c})^\dag } {\langle \hat c^\dag \hat c \rangle_t } \,.
\end{align}
\end{itemize}
We can now write the evolution in terms of the Poissonian increment $dN_t$ that still satisfies $\mathbbm{E}\left[dN_t\right] = \kappa \Tr \left[ \varrho^{\text{(c)}}(t) \hat c^\dag \hat c \right] dt$, obtaining the stochastic feedback master equation for a generic trajectory (including also a system Hamiltonian $\hat{H}$) as
\begin{multline}
d\varrho^\text{(c,fb)} = -i \left[ \hat H, \varrho^\text{(c,fb)} \right]\, dt - \frac{\kappa}{2} \mathcal{H}[\hat c^\dag \hat c] \varrho^\text{(c,fb)} dt  \\
+ \left( \frac{(e^{-i \hat F} \hat c) \varrho^\text{(c,fb)} (e^{-i \hat F} \hat c)^\dag }{\Tr \left[\hat c \varrho^\text{(c,fb)} \hat c^\dag \right]} - \varrho^\text{(c,fb)} \right) d N_t \,. \label{eq:SME_PD_feedack}
\end{multline}
As anticipated above, when we implement a feedback scheme, we are more interested in the unconditional evolution, i.e. on the average state obtained thanks to the implemented feedback scheme.
It is easy to check, by exploiting the property \eqref{eq:PoissonProp}, that in this case the Markovian feedback master equation reads
\begin{align}
\frac{d\varrho^\text{(fb)}}{dt} = -i \left[ \hat H, \varrho^\text{(fb)} \right] + \kappa \mathcal{D} [ e^{-i \hat{F} } \hat c] \varrho^\text{(fb)} \,,
\end{align}
that is the effect of feedback is to replace the jump operator as $ \hat c \rightarrow e^{-i \hat{F}} \hat c$.

As an example of application we refer to Ref.~\cite[Sec.~5.4.3]{wiseman2010quantum} where a photo-detection based feedback protocol is proposed to protect Schrödinger cat states from photon loss.

\subsubsection{Markovian feedback for continuous homodyne detection}
In the case of continuous homodyne detection, the measured infinitesimal photocurrent reads (for simplicity we will consider the homodyne phase $\theta=0$)
\begin{align}
dy_t = \sqrt{\eta \kappa} \langle \hat c + \hat c^\dag \rangle_t \, dt + dw_t \,.
\end{align}
We will apply a Markovian feedback, ruled by an Hamiltonian
\begin{align}
\hat H_\text{fb} (t) = \widetilde{I}(t) \hat F \, ,
\end{align}
where, in order to obtain final results that will be \emph{typographically} easier to understand, we have renormalized the detected photocurrent with the square root of the measurement efficiency $\sqrt\eta$, such that $\widetilde{I}(t) = \frac{dy_t}{dt} \frac{1}{\sqrt\eta}$.

By applying the feedback unitary operator $\hat U_\text{fb} = e^{-i \hat{H}_\text{fb} dt}$ after the measurement, the output state reads
\begin{align}
&\varrho^\text{(c,fb)}(t+dt) = \hat U_\text{fb} \cdot \\
& \left( \frac{ \hat M_{dy_t} \varrho^{\text{(c,fb)}}(t) \hat M_{dy_t}^\dag + \kappa (1-\eta) \hat{c} \varrho^{\text{(c,fb)}}(t) \hat{c}^\dag \,dt } { \Tr[ \hat M_{dy_t}\varrho^{\text{(c,fb)}}(t) \hat M_{dy_t}^\dag +  \kappa (1-\eta)  \hat{c} \varrho^{\text{(c,fb)}}(t) \hat{c}^\dag \,dt ]} \right) \hat U_\text{fb}^\dag \, . \nonumber
\label{eq:KrausFB_homo}
\end{align}
The feedback operator can be expanded up to order $O(dt)$ by exploiting Itô calculus ($dw_t^2 = dt$), obtaining
\begin{align}
\hat U_\text{fb} &= e^{-i \hat{H}_\text{fb} dt} = e^{-i \hat{F} dy_t / \sqrt{\eta}} \nonumber \\
&\approx \mathbbm{1} - i \hat F \left( \sqrt{\kappa} \langle \hat c + \hat c^\dag \rangle_t \, dt + \frac{dw_t}{\sqrt\eta} \right) - \frac{\hat F^2 }{2 \eta} \,dt \,.
\end{align}
By multiplying all the terms in Eq. \eqref{eq:KrausFB_homo},
\begin{align}
&\varrho^\text{(c,fb)}(t+dt) = \left[ \mathbbm{1} - i \hat F \left( \sqrt{\kappa} \langle \hat c + \hat c^\dag \rangle_t \, dt + \frac{dw_t}{\sqrt\eta} \right) - \frac{\hat F^2 }{2 \eta} \,dt  \right] \nonumber \\
& \, \cdot \left[ \varrho^\text{(c,fb)}(t) +  \kappa \mathcal{D}\left[ \hat c \right] \varrho^\text{(c,fb)}(t) dt + \sqrt{\eta\kappa} \mathcal{H}[ \hat c ] \varrho^\text{(c,fb)}(t) \,dw_t \right] \nonumber \\
&\, \quad \cdot  \left[ \mathbbm{1} + i \hat F \left( \sqrt{\kappa} \langle \hat c + \hat c^\dag \rangle_t \, dt + \frac{dw_t}{\sqrt\eta} \right) - \frac{\hat F^2 }{2 \eta} \,dt  \right]
\end{align}
and discarding all the terms of order $o(dt)$ (e.g. all the terms multiplied by $dt^2$ or $dt \,dw_t$), one obtains the stochastic feedback master equation
\begin{multline}
d\varrho^\text{(c,fb)} = \kappa \mathcal{D}[\hat c] \varrho^\text{(c,fb)} \,dt - i \sqrt{\kappa} [ \hat F, \hat c \varrho^\text{(c,fb)} + \varrho^\text{(c,fb)} \hat c^\dag ] \,dt  \\
+  \frac{1}{\eta} \mathcal{D}[\hat F] \varrho^\text{(c,fb)} \,dt + \sqrt{\eta \kappa} \mathcal{H}[\hat c ]\varrho^\text{(c,fb)} \,dw_t - i [\hat F, \varrho^\text{(c,fb)}] \, dw_t \,.
\end{multline}
Notice that if one assumes perfect detection ($\eta = 1$), this last equation can be rewritten as
\begin{multline}
d\varrho^\text{(c,fb)} = - i \sqrt{\kappa} \left[ \frac{\hat c^\dag \hat F + \hat F \hat c}{2} , \varrho^\text{(c,fb)} \right] + \mathcal{D}[\sqrt{\kappa} \hat c - i \hat F] \varrho^\text{(c,fb)}  \\
 + \mathcal{H}[\sqrt{\kappa} \hat c - i \hat F ]\varrho^\text{(c,fb)} \,dw_t \,,
\end{multline}
that is one obtains that each trajectory is described by a \emph{homodyne} SME, with a modified jump operator $\hat{\bar{c}} = \sqrt{\kappa} \hat c - i \hat F$, and with an extra {\em Hamiltonian} term in the dynamics.

If we now focus on the unconditional dynamics, we can easily find the Markovian feedback master equation by using the property $\mathbbm{E}[dw_t] = 0$ and obtaining
\begin{align}
\frac{d \varrho^\text{(fb)}}{dt} = \kappa \mathcal{D}[\hat c] \varrho^\text{(fb)} - i \sqrt{\kappa}[\hat F, \hat c \varrho^\text{(fb)} + \varrho^\text{(fb)} \hat c^\dag ] + \frac{1}{\eta} \mathcal{D}[\hat F] \varrho^\text{(fb)} \,,
\end{align}
where the second term describes the desired effect of feedback, while the last term describes the added noise (due to the white noise in the photocurrent), entering into the system on the operator \emph{conjugated} to the feedback operator $\hat F$. As it is apparent this added noise increases by decreasing the measurement efficiency $\eta$.

The whole evolution is Markovian, and thus we are also able to rewrite it in a \emph{Lindblad form}, i.e.
\begin{multline}
\frac{d \varrho^\text{(fb)}}{dt} =  - i \sqrt{\kappa} \left[ \frac{\hat c^\dag \hat F + \hat F \hat c}{2} , \varrho^\text{(fb)} \right] + \mathcal{D}[\sqrt{\kappa} \hat c - i \hat F] \varrho^\text{(fb)} \\
 + \frac{1-\eta}{\eta} \mathcal{D}[\hat F] \varrho^\text{(fb)} \,.
\end{multline}

We refer to Ref.~\cite{Wang2001} for a simple but very instructive example of homodyne-based feedback protocol for the deterministic preparation of qubit pure state.
\section{Markovian and state-based feedback for Gaussian systems}\label{sec:Gaussian}

In this section we will discuss state-based and Markovian feedback schemes for continuously monitored quantum systems made up of harmonic oscillators.
In particular, we focus on so-called Gaussian regime, i.e. systems obeying a linear Heisenberg-picture dynamics, subject to Gaussian measurements and initialized in Gaussian initial states.
This is a paradigmatic setup~\cite{Serafini2023,Genoni2016} for which many classical results can be often adapted~\cite{Nurdin2017}.
We will not present the derivations of ME and SMEs; rather, we aim to present a treatment that complements Ref.~\cite{Genoni2016}, by discussing measurement-based feedback in the Gaussian scenario.
A similar treatment appears in Ref.~\cite{Serafini2023}.

We will adopt the same notation introduced in Refs.~\cite{Genoni2016,Serafini2023}: a set of $n$ quantum harmonic oscillators is described by a vector of operators 
$$\hat{\bf r}^{\sf T} = ( \hat{q}_1, \hat{p}_1, \dots , \hat{q}_n, \hat{p}_n),$$ 
satisfying the canonical commutation relation $[ \hat{\bf r}, \hat{\bf r}^{\sf T}] = i \Omega$, with $\Omega = i \bigoplus_{j=1}^n \sigma_y $ being the symplectic form\footnote{For more details on the \emph{outer product notation} and on the symplectic form, we suggest to look at Refs.~\cite{Genoni2016,Serafini2023}.}($\sigma_y$ is the $y$ Pauli matrix).
Gaussian states $\varrho$ are equivalently described by the first moment vector $\bar{\bf r} = \Tr[\varrho \hat{\bf r}]$, and covariance matrix
\be
\sigmaCM = \Tr[\varrho \{ (\hat{\bf r} - \bar{\bf r}), (\hat{\bf r} - \bar{\bf r})^{\sf T} \} ] \,.
\ee
We consider a set of quantum harmonic oscillators, characterized by a quadratic Hamiltonian, and linearly interacting with a Markovian Gaussian environment. The corresponding unconditional evolution, that is obtained whenever the environment is not monitored after the interaction with the system, is described by the following equation for first and second moments of the unconditional state $\varrho_{\sf unc}$,
\begin{align}
\frac{d\bar{\bf r}_{\sf unc}}{dt} &= A \bar{\bf r}_{\sf unc} \,, \label{eq:runc} \\
\frac{d \sigmaCM_{\sf unc}}{dt} &= A \sigmaCM_{\sf unc} + \sigmaCM_{\sf unc} A^{\sf T} + D \,, \label{eq:sigmaunc}
\end{align}
where the drift matrix $A$ depends on the system Hamiltonian and on the interaction between system and environment, while the diffusion matrix $D$ depends both on the kind of interaction with the environment, and on the properties of the environment itself (for example its temperature).

If the environment is continuously monitored with Gaussian measurements, i.e. the so-called general-dyne detection, one obtains a stochastic evolution for the first moments of the conditional state $\varrho^{\text{(c)}}$, and a deterministic evolution for its covariance matrix, i.e.
\begin{align}
d\bar{\bf r}_c &= A \bar{\bf r}_c + (E - \sigmaCM_c B) \frac{{\bf dw}}{\sqrt{2}} \,, \label{eq:rc}  \\
\frac{d \sigmaCM_c}{dt} &= A \sigmaCM_c + \sigmaCM_c A^{\sf T} + D - (E - \sigmaCM_c B)(E - \sigmaCM B)^{\sf T} \,, \label{eq:sigmac}
\end{align}
where the matrices $E$ and $B$ depends on the properties of the environment, on the interaction between system and environment, and on the kind of measurement performed, while ${\bf dw}$ is a vector of Wiener increments, satisfying $dw_j dw_k = \delta_{jk} dt$, or in a more compact form $\{ {\bf dw}, {\bf dw}^{\sf T} \}/2=\mathbbm{1}\, dt$. This monitoring yields as a measurement results an infinitesimal continuous photo-current ${\bf dy}_t$, that carries information on the system first moment vector as 
\be
{\bf dy}_t = - \sqrt{2} B^{\sf T} \bar{\bf r}_c \,dt + {\bf dw} \,.
\label{eq:photocurrent}
\ee
We refer again to Refs.~\cite{Serafini2023,Genoni2016} for more details on the matrices appearing in these equations and for more details on the derivation of the equations themselves.

\subsubsection{Unconditional dynamics from conditional dynamics}
While Eqs. \eqref{eq:runc} and \eqref{eq:sigmaunc} are typically derived before introducing the continuous measurement on the environment, 
it is also interesting and useful to show how to obtain them from the Eqs. of the conditional state \eqref{eq:rc} and \eqref{eq:sigmac}.
We remind the reader that the unconditional state is obtained from the conditional one, by averaging over all the trajectories, in formula $\varrho_{\sf unc} = \EE[\varrho^{\text{(c)}}]$.
While it is straightforward to (re)derive Eq.~\eqref{eq:runc} from Eq.~\eqref{eq:rc}, by simply using the property $\EE[{\bf dw}]=0$, the same argument cannot be used for the covariance matrix.
By slightly changing the notation, we write the covariance matrix of conditional and unconditional states respectively as
\begin{align}
\sigmaCM_c &= \langle \{ \hat {\bf r}, \hat{\bf r}^{\sf T} \} \rangle_c - \{ \langle \hat{\bf r} \rangle_c , \langle \hat{\bf r} \rangle_c^{\sf T} \} \,, \\
\sigmaCM_{\sf unc} &= \EE \left[ \langle \{ \hat {\bf r}, \hat{\bf r}^{\sf T} \} \rangle_c  \right] - \{ \EE[\langle \hat{\bf r} \rangle_c] , \EE[\langle \hat{\bf r} \rangle_c^{\sf T} ] \} \,,
\end{align}
where we introduced the notation $\langle \hat A \rangle_c = \Tr[\varrho^{\text{(c)}} \hat A ]$.
It is clear from the formulas above that 
\begin{align}
\sigmaCM_{\sf unc} \neq \EE [ \sigmaCM_c ] \,.
\end{align}
However, one can write
\begin{align}
\sigmaCM_{\sf unc} &= \EE [ \sigmaCM_c ] + \EE[ \{ \langle \hat{\bf r} \rangle_c , \langle \hat{\bf r} \rangle_c^{\sf T} \} ] -  \{ \EE[\langle \hat{\bf r} \rangle_c] , \EE[\langle \hat{\bf r} \rangle_c^{\sf T} ] \} \,. \\
&= \sigmaCM_c + \boldsymbol{\Sigma} \,
\label{eq:sigmaunc2}
\end{align}
where we have exploited the deterministic evolution of $\sigmaCM_c$ and we have defined the excess noise matrix $\boldsymbol{\Sigma} =  \EE[ \{ \langle \hat{\bf r} \rangle_c , \langle \hat{\bf r} \rangle_c^{\sf T} \} ] -  \{ \EE[\langle \hat{\bf r} \rangle_c] , \EE[\langle \hat{\bf r} \rangle_c^{\sf T} ] \}$. 

One can thus derive the variation of the unconditional covariance matrix by writing
\begin{equation}
   \begin{split}
d\sigmaCM_{\sf unc} =  d \sigmaCM_c  + d \boldsymbol{\Sigma} \,. \label{eq:dsigmaunc}
   \end{split}
\end{equation}
The evolution of the excess noise matrix can then be obtained via the equation 
\begin{align}
d\boldsymbol{\Sigma} &=  d\left( \EE[ \{ \langle \hat{\bf r} \rangle_c , \langle \hat{\bf r} \rangle_c^{\sf T} \} ] \right) 
- d \{ \EE[\langle \hat{\bf r} \rangle_c] , \EE[\langle \hat{\bf r} \rangle_c^{\sf T} ] \}  \,,  
\end{align}
which, by exploiting Itô's rule for differentiation for the term
\begin{align}
 & d\left( \EE[ \{ \langle \hat{\bf r} \rangle_c , \langle \hat{\bf r} \rangle_c^{\sf T} \} ] \right) = \EE \left[ \{ d \langle \hat{\bf r} \rangle_c , \langle \hat{\bf r} \rangle_c^{\sf T} \} \right] +  \label{eq:excess} \\
 & \quad + \EE \left[ \{ \langle \hat{\bf r} \rangle_c , d \langle \hat{\bf r} \rangle_c^{\sf T} \} \right] + \EE \left[ \{ d \langle \hat{\bf r} \rangle_c , d \langle \hat{\bf r} \rangle_c^{\sf T} \} \right] \nonumber
\end{align}
and the Wiener increment property $\{ {\bf dw}, {\bf dw}^{\sf T} \}/2 = \mathbbm{1}\,dt$, leads to the result
\begin{align}
d\boldsymbol{\Sigma} &= 
&= A \boldsymbol{\Sigma} + \boldsymbol{\Sigma} A^{\sf T} + (E - \sigmaCM_c B)(E - \sigmaCM B)^{\sf T} \,,
\label{eq:dSigma}
\end{align}
It is then straightforward to observe how, by also exploiting the conditional covariance matrix evolution in Eq.~\eqref{eq:sigmac}, one finally obtains the unconditional evolution in Eq.~\eqref{eq:sigmaunc}.

\subsubsection{Implementation of linear feedback}
We now assume to implement a linear feedback Hamiltonian,
\be
\hat H_{\sf fb} = - \hat{\bf r}^{\sf T} \Omega F {\bf u}(t) \,.
\label{eq:feedbackHam}
\ee
This Hamiltonian corresponds to a displacement in phase-space where the vector ${\bf u}(t)$ describes the amount of displacement, that is chosen according to the particular feedback strategy implemented. The (fixed) matrix $F$ on the other hand describes the {\em kind} of displacement that one can perform; for instance a full rank $F$ matrix corresponds to the case where one can implement displacement in all the possible direction of phase space; a feedback matrix that is not full-rank indicates a certain limitation in the possible displacement that one can perform. For example, a feedback matrix with a single non-zero element $F_{2,2}=\lambda$, describes a feedback Hamiltonian $\hat H_{\sf fb} = \lambda  u_p(t)  \hat{q}$, i.e. a displacement only in momentum and no displacement in position.

As a consequence, the evolution of the conditional covariance matrix is not modified, while the first moment vector evolution is now ruled by the equation
\be
d{\bf r}_c = A {\bf r}_c \,dt + (E - \sigmaCM_c B)\frac{\bf dw}{\sqrt{2}} + F {\bf u}(t) \,dt  \,.
\label{eq:firstmomentFeedback}
\ee
In the following sections we will describe two possible situations, where the vector ${\bf u}(t)$ is chosen respectively according to a state-based (Bayesian) or a Markovian feedback strategies.
\subsubsection{State-based LQG-control feedback}
We assume that the feedback parameter ${\bf u}(t)$ entering in the first-moments evolution equation \eqref{eq:firstmomentFeedback} depends on the whole history of measurement results (photo-current) ${\bf dy}_s$, with $s<t$. This kind of feedback control strategy corresponds to a {\em state-based (Bayesian) feedback} strategy, since knowing the full stream of outcomes gives a knowledge of the state of the system at time $t$, since it can be indeed obtained from the measurement outcomes via a Bayesian update.
The typical aim of a control strategy is to minimize a desired {\em cost} function. Here we will focus on {\em quadratic} cost function, defined as 
\begin{align}
h(t) = \langle \hat{\bf r}^{\sf T} P \hat{\bf r} \rangle_c + {\bf u}^{\sf T} Q {\bf u} \label{eq:costfunction}
\end{align}
where $P$ and $Q$ are two matrices satisfying $P \geq 0$ and $Q > 0$. The first term represents the particular property of the system that we want to minimize, for example $P=\mathbbm{1}_2$ corresponds to minimize the energy of the quantum oscillator, while a matrix with a single non-zero element $P_{1,1}=1$ corresponds to minimize the variance of the quadrature $\hat{q}$ and thus to optimize the corresponding squeezing property. The second term on the other hand quantifies the cost of the linear driving ${\bf u}(t)$ that we are implementing with our feedback strategy. The limit $Q \rightarrow 0$ corresponds to the ideal scenario where displacement in phase-space has zero cost.
Given Eq. \eqref{eq:firstmomentFeedback} and this kind of cost function, we are dealing with the paradigm of {\em LQG-control}, where {\em LQG} stands for {\em Linear} system, {\em Quadratic} cost function and {\em Gaussian} noise. This is indeed a well-known classical control problem~\cite{wiseman2010quantum}; the first main result that has been demonstrated is that the only property of the conditional state that is required to implement the optimal feedback, and thus minimize the cost function $h(t)$, is the first moment vector $\bar{\bf r}_c$. Furthermore, one obtains that ${\bf u}(t)$ depends linearly on $\bar{\bf r}_c$, i.e. one can write
\be
{\bf u}(t) = - K(t) \bar{\bf r}_c \,,
\ee
such that the evolution of the first moments can be rewritten as
\be
d{\bf r}_c = (A - F K(t) ) {\bf r}_c \,dt + (E - \sigmaCM_c B)\frac{\bf dw}{\sqrt{2}} \,.
\ee
Here we will focus on minimizing the chosen quadratic cost function at steady-state, i.e. 
\be
h_{\sf ss} = \lim_{t \to \infty} h(t) \,.
\ee
In this instance one can prove that the optimal matrix $K_{\sf opt}$ reads
\be
K_{\sf opt} = Q^{-1} F^{\sf T} Y \,
\label{eq:Kopt}
\ee
where $Y$ is the solution of the (homogeneous) Riccati equation
\be
A^{\sf T} Y + Y A + P  - Y F Q^{-1} F^{\sf T} Y = 0 \,.
\label{eq:ypsilon}
\ee
By using Eq. \eqref{eq:sigmaunc2}, we can write
\begin{align}
\sigmaCM_{\sf unc}^{\sf ss} &= \sigmaCM_c^{\sf ss} + \boldsymbol{\Sigma}^{\sf ss} \,,
\end{align}
where $\sigmaCM_c^{\sf ss}$ is the deterministic steady-state solution of Eq. \eqref{eq:sigmac}, and $\boldsymbol{\Sigma}^{\sf ss}$ denotes the steady-state excess noise matrix, that is now obtained unconditionally after the feedback protocol. Assuming that the new drift matrix $A - K_{\sf opt}F$ is Hurwitz (i.e. that the system is stable, and the unconditional first moments goes to zero at steady state), the excess noise matrix is simply equal to $\boldsymbol{\Sigma}^{\sf ss} = \mathbbm{E}[ \{ \langle \hat{\bf r} \rangle_c , \langle \hat{\bf r}^{\sf T} \rangle_c \} ]$. By exploiting Eq. \eqref{eq:excess} and the first moment evolution equation \eqref{eq:firstmomentFeedback} (where $\sigmaCM_c$ is replaced by the solution $\sigmaCM_c^{\sf ss}$), one has that, in the large time limit, the evolution of the excess noise matrix reads
\begin{align}
&d \boldsymbol{\Sigma} = d ( \mathbbm{E}[ \{ \langle \hat{\bf r} \rangle_c , \langle \hat{\bf r}^{\sf T} \rangle_c \} ])  \\
&=  (A -FK_{\sf opt}) \, \mathbbm{E}[ \{ \langle \hat{\bf r} \rangle_c , \langle \hat{\bf r}^{\sf T} \rangle_c \} ] \,dt \nonumber \\
& + \mathbbm{E}[ \{ \langle \hat{\bf r} \rangle_c , \langle \hat{\bf r}^{\sf T} \rangle_c \} ] \, (A - F K_{\sf opt}) \,dt + L \left( \mathbbm{E}[ \{{\bf dw}, {\bf dw}^{\sf T} \} ] \right) L^{\sf T}, \nonumber 
\end{align}
where $L=(E - \sigmaCM_c^{\sf ss}B)/\sqrt{2}$, leading to the evolution equation
\begin{multline}
   \label{eq:excessnoise}
   \frac{d \boldsymbol{\Sigma}}{dt} =   (A -F K_{\sf opt}) \boldsymbol{\Sigma} + \boldsymbol{\Sigma} (A - F K_{\sf opt})^{\sf T} \\
+ (E - \sigmaCM_c^{\sf ss} B) (E - \sigmaCM_c^{\sf ss}B)^{\sf T}.
\end{multline}
By solving this equation at steady-state, that is the corresponding Lyapunov equation, one can thus calculate the steady-state excess noise matrix $\boldsymbol{\Sigma}^{\sf ss}$ and assess the performance of the feedback strategy.

It is important to point out again that the optimal feedback strategy, described by the matrix $K_{\sf opt}$, depends on the feedback matrix $F$ as in Eq. \eqref{eq:Kopt}. The beauty of the LQG-control formalism lies in the fact that we are assured that we are getting the best possible result, in terms of the cost function $h_{\sf ss}$, both in the ideal case, where $F$ is a full-rank matrix and thus one can perform displacement in any direction, but also in situations where, for some physical constraints, one can perform displacement only in a given direction of phase-space ( for example when the feedback Hamiltonian reads $\hat{H}_{\sf fb} = \lambda \,u_p(t) \hat{q} $, corresponding to a feedback matrix with a single non-zero element $F_{2,2}=\lambda$).
\subsubsection{Markovian linear feedback}
We now consider the case of Markovian feedback, i.e. we consider the linear displacement in the feedback Hamiltonian \eqref{eq:feedbackHam} depending on the photocurrent, as
\be
{\bf u}(t) = M \, {\bf I} (t) \,,
\ee
where ${\bf I}(t) =  \frac{{\bf dy}_t }{dt}$ and ${\bf dy}_t$ depends linearly on the first moments vector as in Eq. \eqref{eq:photocurrent}.
The evolution of the first moments vector can thus be written as
\begin{align}
d\bar{\bf r}_c &= A \bar{\bf r}_c + (E - \sigmaCM_c B) \frac{{\bf dw}}{\sqrt{2}} + FM \,{\bf dy}_t \\
&=  (A - \sqrt{2} FMB^{\sf T} ) \bar{\bf r}_c + \left(\frac{E - \sigmaCM_c B}{\sqrt{2}} + FM \right){\bf dw} \,.
\end{align}
The goal of feedback is to remove, at least at steady-state, the stochasticity present in the first moments evolution; by looking at the equation above, it is clear that this can be obtained by solving the equation,
\be
E - \sigmaCM_c^{(ss)} B + \sqrt{2} FM = 0 \,,
\ee
leading to an optimal matrix $M_{\sf opt}$ 
\be
M_{\sf opt} = -  \frac{F^{-1} (E - \sigmaCM_c^{\sf ss} B)}{\sqrt{2}} \,.
\label{eq:markovianFB}
\ee
Assuming that the matrix $A^\prime= A - \sqrt{2} FM_{\sf opt}B^{\sf T}$ is {\em Hurwitz}, the first moments vector goes deterministically to zero, and, more importantly, one obtains the {\em optimal} result: an unconditional steady-state covariance matrix equal to the conditional one, i.e. $\sigmaCM_{\sf unc}^{\sf ss}=\sigmaCM_c^{\sf ss}$. We dubbed this result as {\em optimal} as no excess noise matrix is added to the conditional covariance matrix and consequently any quadratic cost function as the one defined in Eq. \eqref{eq:costfunction} with zero cost on the linear driving (i.e. with matrix $Q=0$) is minimized.

As we pointed out in the introduction of Sec.~\ref{s:MarkovianFME}, Markovian feedback is conceptually and experimentally less demanding than state-based feedback as the feedback Hamiltonian only linearly depends on the last measurement outcome ${\bf dy}_t$. In this sense it is quite remarkable that by using only Markovian feedback we obtained the optimal result above, i.e. an unconditional covariance matrix equal to the conditional one (notice that this is another peculiarity of the Gaussian dynamics). However, if we want to make a fair comparison with the LQG-control feedback strategy we have described in the previous subsection, we observe that this result is obtained paying a high price.
\begin{itemize}
\item By looking at Eq. \eqref{eq:markovianFB} it is clear that the optimal Markovian feedback strategy can only be implemented if the feedback matrix $F$ is full-rank (and thus invertible); if displacement in certain directions of phase space are not allowed, this Markovian strategy cannot always be implemented.
\item  We remarked above that this result is optimal in terms of quadratic cost function with zero cost for the linear driving (i.e. for matrix $Q=0$). This reflects the fact that the photocurrent ${\bf I}(t)={\bf dy}_t/ dt$ has unbounded variation, so doing Markovian control is as onerous as doing optimal state-based control with unbounded matrix $K$. In fact even for LQG-control, for a full-rank feedback matrix $F$ and in the limit $Q \rightarrow 0$, one can show that the optimal result $\sigmaCM_{\sf unc}^{\sf ss}=\sigmaCM_c^{\sf ss}$ is equivalently obtained.
\end{itemize}
We will see in the next section an example that will help in understanding the results obtainable via different feedback strategies.
\subsubsection{Example: Squeezing generation in an optical parametric oscillator}
We consider an optical parametric oscillator, i.e. a cavity mode whose interaction with a pumped non-linear crystal is described by the Hamiltonian $\hat{H}_S= -\chi (\hat{q} \hat{p} + \hat{p}\hat{q})/2$. The noisy evolution is ruled by the master equation
\begin{align}
\frac{d\varrho}{dt} = - i [\hat{H}_S,\varrho] + \kappa \mathcal{D}[a]\varrho \,,
\end{align}
where $\kappa$ describes the loss rate of the cavity.\\

\noindent
\uline{Unconditional evolution}\\
The unconditional evolution is Gaussian and can be described by Eqs. \eqref{eq:runc} and \eqref{eq:sigmaunc}, with matrices (see Sec. 6.1 of Ref.~\cite{Genoni2016}): 
\begin{align}
A= 
\left(
\begin{array}{c c}
- (\chi + \kappa/2) & 0 \\
0 & \chi - \kappa/2 
\end{array}
\right)\,, \,\,\,\,\,  
D= \kappa \mathbbm{1}_2 \,.
\end{align}
In the following we will restrict to stable systems, and thus we will assume that the drift matrix $A$ is Hurwitz, i.e. $|\chi | < \kappa/2$ (for simplicity we will also assume positive coupling constant $\chi>0$). The steady-state unconditional covariance matrix can be readily obtained, 
\be
\sigmaCM_{\sf unc}^{\sf ss} = \left(
\begin{array}{c c}
\frac{\kappa}{\kappa+2 \chi} & 0 \\
0 & \frac{\kappa}{\kappa - 2 \chi} 
\end{array}
\right)
\ee
It is evident that squeezing in the $\hat{q}$ quadrature, i.e. $\langle \Delta \hat{q}\rangle^2 = \sigma_{11}/2 < 1/2$, is always obtained for any value of $0<\chi<\kappa/2$. In particular, in the limit of instability ($\chi \rightarrow \kappa/2$), one gets as maximum squeezing achievable $\langle \Delta \hat{q}\rangle^2 = 1/4$, corresponding to a {\em 3dB squeezing limit}\footnote{Squeezing is often quantified, in decibels, as  $dB = 10 \log_{10} \frac{\langle \Delta \hat{q}\rangle^2}{ \langle \Delta \hat{q}\rangle_0^2}$, where $\langle \Delta \hat{q}\rangle_0^2 = 1/2$ denotes the vacuum variance. In this instance, one has $dB = 10 \log_{10} 2 \approx 3.01$.}.\\

\noindent
\uline{Conditional evolution via homodyne detection --- Optimal Markovian feedback}

We now assume that a continuous monitoring of the $\hat{q}$ quadrature is performed via continuous measurement of the output modes with efficiency $\eta$.
This corresponds to a SME
\be
d\varrho^{\text{(c)}} = -i  [ \hat{H}_s ,\varrho^{\text{(c)}} ] \,dt + \kappa \mathcal{D}[a]\varrho^{\text{(c)}}  \,dt + \sqrt{\eta\kappa} \mathcal{H}[a]\varrho^{\text{(c)}}  \,dw \,,
\ee
leading to a continuous photocurrent $dy_t = \sqrt{\eta \kappa} \Tr[\varrho^{\text{(c)}} (a + a^\dag)]\,dt + dw$. Also in this case the dynamics is Gaussian and can be described via Eqs. \eqref{eq:rc} and \eqref{eq:sigmac}, where the matrices $B$ and $E$ read
\be
B = E =  \left(
\begin{array}{c c}
-\sqrt{\eta \kappa} & 0 \\
0 & 0 
\end{array}
\right) \,.
\ee
In the following we will restrict to the ideal case of perfect monitoring $\eta=1$. In detail, one shows that the first moment evolution can be written as
\begin{align}
d \langle \hat{q} \rangle_c &= - (\chi + \kappa/2)  \langle \hat{q} \rangle_c \,dt+ \sqrt{\frac{\kappa}{2}} \left( 2 \langle \Delta \hat{q}^2 \rangle_c -1 \right) \, dw \,,\\
d \langle \hat{p} \rangle_c &= - (\kappa/2 - \chi)  \langle \hat{p} \rangle_c \,dt+  \,,
\end{align}
showing how the stochastic part enters only in the evolution of the monitored $\hat{q}$ quadrature. \\
Also in this case the steady-state covariance matrix can be obtained analytically, yielding
\be
\sigmaCM_c^{\sf ss} = \left(
\begin{array}{c c}
\frac{\kappa - 2 \chi}{\kappa} & 0 \\
0 & \frac{\kappa}{\kappa - 2 \chi} 
\end{array}
\right) \,.
\ee
This clearly corresponds to an enhanced squeezing in the $\hat{q}$ quadrature respect to the unconditional result for any value of $\chi$. In particular near instability, i.e. for $\chi \rightarrow \kappa/2$, one gets infinite squeezing as the variance goes to zero.\\
Remarkably, this result can be obtained unconditionally with first moments equal to zero, by implementing the optimal Markovian feedback strategy we discussed above. We remind that this implies a full-rank feedback matrix, e.g. $F= \lambda \mathbbm{1}_2$, and unbounded linear displacement generated by the corresponding feedback Hamiltonian. In  turn the optimal Markovian matrix has a single non-zero element $(M_{\sf opt})_{1,1} = (\chi / \lambda) \sqrt{2/\kappa} $, such that the corresponding feedback Hamiltonian, linear in the instantaneous photocurrent $I(t)= dy_t / dt$, reads
\be
\hat{H}_{\sf fb} = \left(\chi \sqrt{\frac{2}\kappa} \frac{dy_t}{dt} \right)  \hat{p}  \,.
\ee
As expected the generator of the optimal feedback Hamiltonian is the momentum operator $\hat{p}$, as the goal is to remove the stochastic part in the evolution of the $\hat{q}$ quadrature. For this reason, in this specific example, one could get the optimal result via Markovian feedback, also with a non-full rank $F$ matrix, e.g.
\be
F =  \left(
\begin{array}{c c}
\lambda & 0 \\
0 & 0
\end{array}
\right) \,.
\label{eq:nonfullrankF}
\ee
\\ 
\noindent
\uline{State-based LQG-control}:\\
We will now address the same problem, but we will implement the optimal state-based LQG-control feedback previously described in this section.
As we want to maximize squeezing of the $\hat{q}$ quadrature (i.e. to reduce its variance), the cost function that we want to minimize is obtained via Eq.~\eqref{eq:costfunction} by choosing the matrices 
\begin{align}
P = 
\left(
\begin{array}{c c}
1 & 0 \\
0 & 0
\end{array}
\right) \,,  \,\,\,\,\, Q = q \mathbbm{1}_2 \,.
\end{align}
We are thus giving the same {\em cost} for displacement along any quadrature in phase-space, while the parameter $q$ is a weighting constant balancing the two contributions to the cost function. If the magnitude of linear driving is relatively unimportant, the parameter $q$ can be chosen small, with the limit $q\rightarrow 0$ if we allow for unbounded displacement.

We will first consider the case where the feedback matrix $F$ is full rank, for simplicity $F=\lambda \mathbbm{1}_2$, i.e. one can implement displacement along any direction in phase-space. The matrices $Y$ and $K_{\sf opt}$ can be readily obtained via Eqs. \eqref{eq:ypsilon} and \eqref{eq:Kopt} (we will not report here all the analytical results as some of them are quite cumbersome) and, consequently the steady-state excess noise matrix $\boldsymbol{\Sigma}^{\sf ss}$ via Eq.~\eqref{eq:excessnoise}.
As a result, one obtains the unconditional covariance matrix
\begin{align}
\sigmaCM_{\sf unc}^{\sf ss} = \sigmaCM_{\sf c}^{\sf ss} + 
\left(
\begin{array}{c c}
f_A ( \chi, \kappa, \lambda, q) & 0 \\
0 & 0
\end{array}
\right),  \nonumber \\
 \textrm{with} \,\,\, f_A(\chi, \kappa,\lambda, q) = \frac{4 q \chi^2}{\kappa \sqrt{q (4\lambda^2 +q(\kappa + 2 \chi)^2)}}
\,.
\end{align}
We thus observe added noise on the unconditional variance of the $\hat{q}$ quadrature, represented by the function $f_A(\chi,\kappa,q)$. As expected, this function is monotonically increasing with $q$ (larger cost of displacement, implies larger noise on the unconditional steady-state) and  we have
\be
\lim_{q \rightarrow 0} f_A(\chi,\kappa, \lambda, q) = 0 \,,
\ee
that is, for zero cost of the displacement, one gets the optimal result $\sigmaCM_{\sf unc}^{\sf ss}=\sigmaCM_{\sf c}^{\sf ss}$.

It is easy to check that the same result can in fact obtained if we consider a feedback matrix $F$ that is not full rank, but that allows displacement only along the $\hat{q}$ quadrature, that is as the one described in Eq. \eqref{eq:nonfullrankF}. In fact, as we pointed out before, the stochastic part in the first moment evolution is present in the evolution of the $\hat{q}$ quadrature only. Analogously, if we apply LQG-feedback control, but with a feedback matrix
\be
F =  \left(
\begin{array}{c c}
0 & 0 \\
0 & \lambda
\end{array}
\right) \,.
\ee
that corresponds to Hamiltonian of the form $\hat{H}_{\sf fb} = \lambda \, u_q(t) \hat{q} $, and thus allows only displacement only along $\hat{p}$, the feedback operation is completely useless, one obtains $K_{\sf opt}=0$, and one gets the unconditional covariance matrix
\be
\sigmaCM_{\sf unc}^{\sf ss} = \left(
\begin{array}{c c}
\frac{\kappa}{\kappa+2 \chi} & 0 \\
0 & \frac{\kappa}{\kappa - 2 \chi} 
\end{array}
\right) \,,
\ee
as in the no-monitoring/no-feedback scenario. \\
Finally we will consider a feedback matrix $F$ not full rank, but of the form
\be
F = \lambda  \left(
\begin{array}{c c}
1 & 1 \\
0 & 0
\end{array}
\right) \,,
\ee
corresponding to a feedback Hamiltonian of the form $\hat{H}_{\sf fb} = \lambda \,u(t) (\hat{q} + \hat{p})$, and thus leading to displacement along the conjugated quadrature $\hat{x}_{-\pi/4} = (\hat{q} - \hat{p})/\sqrt{2}$. In this case the unconditional covariance matrix reads 
\begin{align}
\sigmaCM_{\sf unc}^{\sf ss} &= \sigmaCM_{\sf c}^{\sf ss} + 
\left(
\begin{array}{c c}
f_B ( \chi, \kappa, \lambda, q) & 0 \\
0 & 0
\end{array}
\right) , \\
\textrm{with}& \,\,\, f_B(\chi, \kappa, \lambda, q) = \frac{8 q \chi^2}{q(\kappa+2 \chi)+ \sqrt{q (8+q(\kappa + 2 \chi)^2)}} \nonumber
\,.
\end{align}
As above we thus obtain added noise on the $\hat{q}$ quadrature, represented by the function $f_B(\chi,\kappa,\lambda,q)$. One can check that this function is monotonically increasing with $q$ and that 
$$
f_B(\chi,\kappa,\lambda,q) \geq f_A(\chi,\kappa,\lambda,q),
$$
that is in general the added noise in this scenario is always larger than the one obtainable in the ideal scenario of full-rank feedback matrix: clearly, in this case one can still displace along $\hat{q}$, but with the same amount of driving, i.e. at fixed cost $q$, the displacement will be lower than the one obtainable in the ideal case. However one still obtains
\be
\lim_{q \rightarrow 0} f_B(\chi,\kappa, \lambda, q) = 0 \,,
\ee
that is for zero driving cost, one can still obtain the optimal result of zero excess noise matrix.

\section{Outlooks}\label{sec:outlooks}

In this introductory notes we have simply started to scratch the surface of a very vast topic that has many ramifications.
We mention here a few theoretical research directions that we regard as topical, without the pretense of being exhaustive nor unbiased.

In the main text we have presented derivations only for an environment initially in the vacuum state, this is most often the scenario of a quantum system playing the role of an \emph{emitter} that spontaneously emits electromagnetic radiation which can be detected, i.e. the setup of fluorescence~\cite{Lewalle2019}.
The same approach can be extended to other temporally uncorrelated initial states, such as thermal (Gibbs) states~\cite{Genoni2014a,Wiseman1994} and broadband squeezed vacuum~\cite{Dabrowska2016,wiseman2010quantum}, or coherent states (this is equivalent to a classical driving on the quantum system~\cite{wiseman2010quantum}).
We provide a brief account of these scenarios in Appendix~\ref{s:genericbath}.
However, many of these ideas can be taken beyond the strict Markovian white noise formalism we have employed here, and more general initial states can be dealt with, at the expense of more complicated dynamical equations, such as Fock~\cite{Baragiola2012,Baragiola2017,Dabrowska2019} and finite-bandwith squeezed states~\cite{Gross2022}.
To the best of our knowledge, the effect of feedback strategies in these more general scenarios has not been investigated yet.

From the point of view of concrete applications, the interplay of measurement-based feedback with machine learning techniques, as recently studied in~\cite{Borah2021,Fallani2022,Porotti2022}, will certainly become more important in the future.
As mentioned in the introduction, continuous monitoring is intrinsically related to statistical inference problems, indeed the idea of using continuously monitored quantum system as \emph{sensors} is one of the most well-developed aspects of the formalism.
In particular, these ideas have been studied both from a general quantum statistics point of view~\cite{Gammelmark2014,Guta2016} and for more concrete problems in the spirit of (noisy) quantum metrology~\cite{Tsang2013a,Albarelli2017a,Albarelli2018a,Rossi2020}, and for real-time tracking of external fields~\cite{Amoros-Binefa2021,khanahmadiTimedependentAtomicMagnetometry2021}.

Techniques based on feedback and continuous measurements have started to be employed also in the context of quantum thermodynamics, e.g. for quantum batteries~\cite{Mitchison2021,Morrone2023}.
We also mention that ideas and concepts useful in the field of continuously-monitored quantum systems general appear in other fields too, since output currents of open quantum systems and their fluctuations are a central object in condensed matter and thermodynamics~\cite{Landi2023}.

Finally, we conclude by mentioning that we have treated here only \emph{measurement-based} feedback, however another approach is possible, i.e. \emph{coherent} feedback, in which the output modes are not measured, but fed back to the input of the system~\cite{zhangQuantumFeedbackTheory2017}.
While coherent feedback seems intuitively more powerful, the comparison between the two approaches requires a nuanced discussion~\cite{Harwood2023}.

\section*{Acknowledgments}
The material in this manuscript derives from the lecture notes for the course ``Coherence and control of quantum systems'' that MGG has been teaching at the University of Milan since 2018; some parts of these lecture notes were adapted from FA PhD thesis~\cite{Albarelli2018e}.

MGG would like to thank all the students that contributed to improve this material over the years. 
We also thank M.~Brunelli, A.~Doherty, M.~S.~Kim, J.~Kołodyński, S.~Mancini, K.~Mølmer, M.~G.~A.~Paris, M.~A.~C.~Rossi, J.~Ralph, P.~Rouchon, A.~Serafini, A.~Smirne and H.~M.~Wiseman for many fruitful and inspiring discussions on these subjects.

FA acknowledges support from Marie Skłodowska-Curie Action EU-HORIZON-MSCA-2021PF-01 (project \mbox{QECANM}, grant n.~701154); MGG acknowledges support from Italian Ministry of Research via the the PRIN 2022 project CONTRABASS (contract n.~2022KB2JJM).

\appendix
\section{Stochastic master equations for more general bath statistics}
\label{s:genericbath}
To simplify the discussion and make our treatment more pedagogical we have only dealt with an initial vacuum state of input modes, i.e. a zero-temperature bath in which the system can emit quanta of radiation.
In this appendix we briefly describe the results obtained for more general statistics of the input modes, but remaining in the assumption of no temporal correlations.
We focus in particular on a squeezed thermal bath in the so-called broadband approximation and on the case of coherent driving.
We will only consider the SMEs pertaining to continuous homodyne detection, since trying to describe photon counting in this ideal scenario of a photodetector with infinite temporal resolution, i.e. infinite bandwidth, is problematic.
As a matter of fact, a detector sees a divergent photon flux for a thermal squeezed input field and does not provide information on the system, see~\cite[Sec.~4.3.3]{wiseman2010quantum}.

\subsection{Generic Gaussian bath}
We start by reminding ourselves the statistics for the operators $\{ \hat{b}_\omega \}$ for a generic Gaussian bath exhibiting squeezing at frequency $\omega_0$ and with zero first moments:
\begin{equation}
\label{eq:bathstatistics1} 
\begin{split}
    \langle \hat{b}_\omega \rangle_{\mathcal{B}} &= 0 \,,  \\
    \langle \{ \hat{b}_\omega, \hat{b}_{\omega^\prime}^\dag \} \rangle_\mathcal{B} &= (2 N + 1 )\delta(\omega - \omega^\prime)  \,,  \\
    \langle \{ \hat{b}_\omega, \hat{b}_{\omega^\prime} \} \rangle_\mathcal{B} &= M \, \delta(2 \omega_0 - \omega - \omega^\prime)  \,.
\end{split}
\end{equation}
Here, the fact that $N$ and $M$ are constants and not functions of $\omega$ reflects the the so-called ``broadband'' approximation, which results into the following white-noise properties for the input operators $\{\hat{b}_t \}$:
\begin{equation}
\label{eq:bathstatistics2} 
\begin{split}
    \langle \hat{b}_t \rangle_{\mathcal{B}} &= 0  \,,  \\
    \langle \{ \hat{b}_t , \hat{b}_{t^\prime}^\dag \} \rangle_\mathcal{B} &= (2 N + 1 ) \delta(t - t^\prime)  \,, \\
    \langle \{ \hat{b}_t , \hat{b}_{t^\prime} \} \rangle_\mathcal{B} 
    &= M \, \delta(t - t^\prime) \, .
\end{split}
\end{equation}
For simplicity we assume $M$ to be real-valued in all the calculations, nonetheless the final results will be given for the more general case of a complex-valued $M$ (the more general derivation follows the same idea).

For non negligible values of $N$ and $M$, i.e. when $\omega_0$ is not large enough to assume the average number of thermal photons $N = N(\omega_0) \approx 0$ or for a squeezed bath ($M \neq 0$), one has to replace the vacuum state in Eq.~\eqref{eq:Born} with a generic squeezed thermal state $\mu_t$, having a diagonal covariance matrix $\boldsymbol{\sigma}={\rm diag}(2N +1 +2M, 2N+1 - 2M)$.
The master equation ruling the unconditional dynamics can then be obtained by following the same procedure outlined in Sec.~\ref{subsec:Lindblad}, obtaining the well-known result~\cite{wiseman2010quantum,Gross2022}
\begin{align}
\frac{d \varrho(t)}{dt} &= \kappa (N+1) \mathcal{D}[ \hat c] \varrho(t) + \kappa N \mathcal{D}[ \hat{c}^\dag] \varrho(t) \nonumber \\ 
&\,\,\, + \frac{\kappa M}{2} [ \hat{c}^\dag, [\hat{c}^\dag,\varrho(t)]] + \frac{\kappa M^*}{2} [ \hat{c}, [\hat{c},\varrho(t)]] .
\end{align}

To derive the SME corresponding to continuous homodyne detection we will
follow the procedure used in Ref. \cite{Wiseman1994}. We start by transforming 
the bath state into a Wigner probability distribution obtaining the operator (in the
system Hilbert space)
\begin{align} 
\widetilde{W}(t) &= \int  \frac{d^2 \lambda }{\pi^2} {\rm Tr}_{\mathcal B} \left[ R\: e^{\left\{\lambda (\hat{B}_t^{\dag}-\alpha^*)-\lambda^*(\hat{B}_t-\alpha) \right\}}\right] \\
&= W_t(\alpha) \varrho(t) .
\end{align}
where 
\begin{widetext}
\begin{align}
W_t(\alpha) = \frac{1}{\pi\sqrt{(N+1/2)^2 - M^2}} \exp\left\{ - \frac{(N+1/2) |\alpha|^2 -\frac{M}{2} (\alpha^2 + \alpha^{* 2}) }{(N+1/2)^2 - M^2} \right\}
\end{align}
\end{widetext}
denotes the Wigner function of the Gaussian state $\mu_t$. The state then evolves according to Eq.~\eqref{eq:evolutionR}, which in the Wigner function picture and up to order $O(\sqrt{\kappa dt})$, reads
\begin{align}
\widetilde{W}(t+dt) &= \widetilde{W}(t) + \sqrt{\kappa dt}\left[ (\alpha^* -\frac12 \partial_\alpha) \hat{c} \widetilde{W}(t) - \right. \nonumber \\
& \: \left. - (\alpha+\frac12 \partial_{\alpha^*})\hat{c}^\dag \widetilde{W}(t) - (\alpha^*+\frac12\partial_\alpha) \widetilde{W}(t) \hat{c} + 
\right. \nonumber \\ 
&\:\: + \left. (\alpha-\frac12 \partial_{\alpha^*}) \widetilde{W}(t) \hat{c}^\dag \right] + O(\kappa dt) \label{eq:wignerevolv} \:.
\end{align}
We now consider the measurement of the operator $\hat{x} = (\hat{B}_t + \hat{B}_t^\dag)/\sqrt{2}$; the unnormalized conditional state in terms of the Wigner function, upon obtaining the measurement result $x$, can then be obtained by integrating $\widetilde{W}(t+dt)$ over the variable $y=-i(\alpha-\alpha^*)$. By performing the derivatives and the integrals, one obtains 
\begin{widetext}
\begin{align}
\widetilde{W}_c(t) &= p_x(t) \left\{ \varrho(t) + \frac{\sqrt{\kappa dt}}{L} x \left[ (N+M+1) \hat{c} \varrho(t) - (N+M) \hat{c}^\dag \varrho(t) + (N+M+1)\varrho(t) \hat{c}^\dag - (N+M) \varrho(t) \hat{c} \right] \right\}+ O(\kappa dt) \,,
\end{align}
\end{widetext}
where $p_x(t)$ is a Gaussian distribution centered in zero and with variance $L=2 N+1 + 2M$. By calculating the trace of the conditional state we obtain the probability of obtaining the result $x$ from the measurement,
\begin{align}
p_x(t+dt) &= \hbox{Tr}[\widetilde{W}_c(t)]  \\
&= p_x(t) \left( 1 + \sqrt{\kappa dt} x L^{-1} \langle \hat{c} + \hat{c}^\dag\rangle_t + O(\kappa dt) \right) \nonumber
\end{align}
that, at the order $O(\sqrt{\kappa dt})$, corresponds to a Gaussian probability distribution centered in $\sqrt{\kappa dt /2} \langle \hat{c} + \hat{c}^\dag\rangle_t$ and with variance $L/2$. We can thus introduce the new stochastic increment $dy_t$ and the corresponding photocurrent $I(t)$ such that
\begin{align}
dy_t &= I(t) dt = \sqrt{2 dt} x  = \sqrt{\kappa} \langle \hat{c} + \hat{c}^\dag\rangle \,dt + \sqrt{L}\, dw_t \,. \label{eq:dwthermal}
\end{align}
After having obtained the normalized conditional state by using the formula
\begin{align}
\varrho^{\text{(c)}} (t+dt) &= \frac{\widetilde{W}_c(t+dt)}{p_x(t+dt)} \:,
\end{align}
and by doing the substitution in (\ref{eq:dwthermal}), one obtains the SME
\begin{widetext}
\begin{align}
d\varrho^{\text{(c)}}(t) &= \varrho^{\text{(c)}} (t+dt)  - \varrho^{\text{(c)}} (t) \nonumber \\
&=  \sqrt{\kappa}\mathcal{H} \left[ \left(N+ M^* + 1\right) \hat{c} - (N+M) \hat{c}^\dagger \right] \varrho^{\text{(c)}}(t) \, \frac{dw_t}{\sqrt{L}} \nonumber \\
& \qquad \qquad \qquad \qquad + \kappa (N+1) \mathcal{D}[ \hat c] \varrho^{\text{(c)}}(t) + \kappa N \mathcal{D}[ \hat{c}^\dag] \varrho^{\text{(c)}}(t) + \frac{\kappa M^*}{2} [ \hat{c}, [\hat{c},\varrho^{\text{(c)}}(t)]] +\frac{\kappa M}{2} [ \hat{c}^\dag, [\hat{c}^\dag,\varrho^{\text{(c)}}(t)]] \, ,
\end{align}
\end{widetext}
where we have added the terms of order $O(\kappa dt)$ from the unconditional master equation.
We remark that in the case of a thermal bath (with no squeezing, $M=0$) one thus obtains
\begin{multline}
d\varrho^{\text{(c)}}(t) 
=  \sqrt{\kappa}\mathcal{H}[(N + 1) \hat{c} - N \hat{c}^\dagger] \varrho^{\text{(c)}}(t) \, \frac{dw_t}{\sqrt{2N+1}} \\
 + \kappa (N+1) \mathcal{D}[ \hat c] \varrho^{\text{(c)}}(t) + \kappa N \mathcal{D}[ \hat{c}^\dag] \varrho^{\text{(c)}}(t)\,,
\end{multline}
with a photocurrent $dy_t= \sqrt{\kappa} \langle \hat{c} + \hat{c}^\dag\rangle_t + \sqrt{2 N + 1}\, dw_t$.

A detailed derivation of the SME in the case of thermal bath for a general-dyne detection can be found in ~\cite{Genoni2014a}; in particular for heterodyne detection one finds
\begin{widetext}
\begin{multline}
d\varrho^{\text{(c)}}(t) 
=  \sqrt{\kappa}\mathcal{H}[(N + 1) \hat{c} - N \hat{c}^\dagger] \varrho^{\text{(c)}}(t) \, \frac{dw_t^{(1)}}{\sqrt{2(N+1)}}  +  \sqrt{\kappa}\mathcal{H}[(N + 1) (i \hat{c}) - N (-i \hat{c}^\dagger)] \varrho^{\text{(c)}}(t) \, \frac{dw_t^{(2)}}{\sqrt{2(N+1)}} \\
+ \kappa (N+1) \mathcal{D}[ \hat c] \varrho^{\text{(c)}}(t) + \kappa N \mathcal{D}[ \hat{c}^\dag] \varrho^{\text{(c)}}(t)\,,
\end{multline}
\end{widetext}
with photocurrents
\begin{equation}
   \begin{split}
d y_t^{(1)} & = \sqrt{\kappa} \langle \hat c + \hat c^\dag \rangle_t dt + \sqrt{2(N+1)}\,dw_t^{(1)} \, ,  \\
d y_t^{(2)} & = \sqrt{\kappa} \langle i (\hat c - \hat c^\dag )\rangle_t dt + \sqrt{2(N+1)}\, dw_t^{(2)} \, .  
   \end{split}
\end{equation}

We now give an alternative derivation in the case of a squeezed vacuum bath, that is when the condition $|M|^2 = N(N+1)$ holds.
In this scenario the input states of all the input modes, i.e. in the usual collision model approach, can be written as
\begin{align}
\mu_t = \hat{S}(r) |0\rangle_t \langle 0| \hat{S}(r)^\dag \,,
\end{align}
where we have introduced the squeezing operator $\hat{S}(r) = \exp\{(r/2) ( \hat{B}_t^{\dag 2} - \hat{B}_t^2)\}$ with $r=\ln(1+2N +2 \sqrt{N(1+N)})/2$, and we have assumed $M\in \mathbbm{R}$. One can then notice that the derivation of unconditional and SMEs can be reduced to the one with an input vacuum state, by considering the modified evolution operator
\begin{align}
    \hat{\widetilde{U}}(t,t+dt) = (\mathbbm{1} \otimes \hat{S}(r)^\dag ) \hat{U}(t,t+dt) (\mathbbm{1} \otimes \hat{S}(r))
\end{align}
with $\hat{U}(t,t+dt)$ as in Eq.~\eqref{eq:input_U_sqrtdt}. The squeezing operator acts on the input operators as $\hat{S}(r)^\dag \hat{B}_t \hat{S}(r)= \mu \hat{B}_t + \nu \hat{B}_t^\dag$, with $\mu=\cosh r$ and $\nu=\sinh r$; consequently one obtains
\begin{widetext}
\begin{align}
    \hat{\widetilde{U}}(t,t+dt) &= \exp \left[ \sqrt{\kappa dt} \left( \hat{c}\otimes (\mu \hat{B}_t^\dag + \nu \hat{B}_t) - \hat{c}^\dag \otimes (\mu \hat{B}_t + \nu \hat{B}_t^\dag) \right) \right] \,,\nonumber \\
    &= \exp \left[ \sqrt{\kappa dt} \left( ( \mu \hat{c} -\nu \hat{c}^\dag )\otimes \hat{B}_t^\dag - (\mu \hat{c}^\dag - \nu \hat{c}) \otimes \hat{B}_t  \right) \right] = \exp \left[ \sqrt{\kappa dt} \left( \hat{\widetilde{c}} \otimes \hat{B}_t^\dag - \hat{\widetilde{c}}^\dag \otimes  \hat{B}_t \right) \right]
\end{align}
\end{widetext}
where have defined the operator $\hat{\widetilde{c}} =  \mu \hat{c} -\nu \hat{c}^\dag$. It is then straightforward to observe that the unconditional and the homodyne SMEs can be be simply obtained from the zero-temperature ones, by simply replacing the jump operator as 
\begin{align}
\frac{d\varrho(t)}{dt} &= \kappa \mathcal{D}[\mu \hat{c} -\nu \hat{c}^\dag] \varrho(t) \,, \\
d\varrho^{\text{(c)}}(t) &= \kappa \mathcal{D}[\mu \hat{c} -\nu \hat{c}^\dag] \varrho^{\text{(c)}}(t) \,dt + \sqrt{\kappa} \mathcal{H}[\mu \hat{c} -\nu \hat{c}^\dag]\varrho^{\text{(c)}}(t) \,dw_t \,,  
\end{align}
where the homodyne photocurrent reads 
\begin{align}
    dy_t = I(t)dt = \sqrt{\kappa} e^{- r}\langle \hat{c} + \hat{c}^\dag \rangle_t \,dt + dw_t \,.
\end{align}
\subsection{Coherent driving}
We now assume a monochromatic driving, corresponding to a zero-temperature bath  with non-zero first moments
\begin{align}
    &\langle \hat{b}_\omega \rangle_{\mathcal{B}} = \beta \, \delta(\omega - \omega_0) \,, \nonumber \\
    &\langle \{ \hat{b}_\omega - \langle \hat{b}_\omega \rangle_{\mathcal{B}}  , \hat{b}_{\omega^\prime}^\dag - \langle \hat{b}^\dag_{\omega^\prime} \rangle_{\mathcal{B}}  \} \rangle_\mathcal{B} = \delta(\omega - \omega^\prime)  \,. \nonumber
\end{align}
This is then translated into the following relations for the input operators
\begin{align}
    &\langle \hat{b}_t \rangle_{\mathcal{B}} = \bar{\beta}  \,, \nonumber \\
    &\langle \{ \hat{b}_t - \bar{\beta} , \hat{b}_{t^\prime}^\dag - \bar\beta^* \} \rangle_\mathcal{B} = \delta(\omega - \omega^\prime)  \, \nonumber \,, 
\end{align}
where $\bar{\beta} = \sqrt{2\pi}\beta$ is in general a complex number and will be considered for simplicity real in what follows. The equations here above tell us that the input state of the collision model can be taken equal to a coherent state $|\beta\rangle_t$, satisfying the property $\hat{B}_t |\beta\rangle_t = (\bar{\beta} \sqrt{dt})|\beta\rangle_t$. Under this assumption the density operator describing system and environment at time $t+dt$ reads
\begin{widetext}
\begin{align}
R(t+dt) &= \varrho(t) \otimes |\beta\rangle\langle \beta| +  \label{eq:globalev_coherent} \\
& + \sqrt{\kappa} \left( \hat c \varrho(t) \otimes \hat{B}_t^\dag|\beta\rangle \langle \beta| -  \hat{c}^\dag \varrho(t) \otimes \hat{B}_t |\beta\rangle \langle \beta| + \mathrm{h.c.} \right) \, \sqrt{dt}  \nonumber \\
& + \kappa \left( \hat c \varrho(t) \hat c^\dag \otimes \hat{B}_t^\dag|\beta\rangle \langle \beta|\hat{B}_t \right) \, dt +  \frac{\kappa}{2} ( \sqrt{2} \hat c^2 \varrho(t) \otimes \hat{B}_t^{\dag 2}|\beta\rangle\langle \beta| - \hat{c}^\dag \hat c \varrho(t) \otimes \hat{B}_t \hat{B}_t^{\dag} |\beta\rangle\langle \beta | + \mathrm{h.c.}) \, dt \,. \nonumber
\end{align}
\end{widetext}
By applying the coherent state property above, and by keeping the terms up to order $O(dt)$, one can then derive the unconditional master equation as in Sec.~\ref{subsec:Lindblad}, yielding
\begin{align}
    \frac{d\varrho(t)}{dt} = -i [\hat{H}_\beta, \varrho(t)] + \kappa \mathcal{D}[\hat{c}]\varrho(t) \,,
\end{align}
where a (driving) Hamiltonian term is now present:
\begin{align}
    \hat{H}_\beta = i \sqrt{\kappa} \bar{\beta} (\hat{c} - \hat{c}^\dag) \,.
\end{align}
Similarly, also in the SME one simply needs to add the driving term, obtaining
\begin{multline}
   d\varrho^{\text{(c)}}(t) = -i [\hat{H}_\beta, \varrho^{\text{(c)}}(t)] \,dt + \kappa \mathcal{D}[\hat{c}]\varrho^{\text{(c)}}(t)\,dt \\
    +\sqrt{\kappa} \mathcal{H}[\hat{c}]\varrho^{\text{(c)}}(t)\, dw_t
\end{multline}
with $dy_t = \sqrt{\kappa}\langle \hat{c} + \hat{c}^\dag \rangle_t \,dt + dw_t$.
\bibliography{biblioContMeasTutorial}

\end{document}